\documentclass[%
 reprint,
 amsmath,amssymb,
 aps,
]{revtex4-2}
\usepackage{makecell}
\usepackage{graphicx}
\usepackage{dcolumn}
\usepackage{bm}
\usepackage{hyperref}
\usepackage{xcolor}
\DeclareUnicodeCharacter{2061}{}
\begin{document}

\preprint{APS/123-QED}

\title{Graph Theorem for Chiral Exact Flat Bands at Charge Neutrality}

\author{Gurjyot Sethi\textsuperscript{$\dagger$,1}}
\author{Bowen Xia\textsuperscript{$\dagger$,1}}
\author{Dongwook Kim\textsuperscript{1}}
\author{Hang Liu\textsuperscript{2,3}}
\author{Xiaoyin Li\textsuperscript{1}}
\author{Feng Liu\textsuperscript{1}}
\email{fliu@eng.utah.edu}
\affiliation{\textsuperscript{1}Department of Materials Science \& Engineering, University of Utah, Salt Lake City, Utah 84112, USA\\ \textsuperscript{2}Songshan Lake Materials Laboratory, Dongguan, Guangdong 523808, People’s Republic of China\\
\textsuperscript{3}Beijing National Laboratory for Condensed Matter Physics and Institute of Physics, Chinese Academy of Sciences, Beijing 100190, People’s Republic of China}

\date{\today}

\begin{abstract}
Chiral exact flat bands (FBs) at charge neutrality have attracted much recent interest, presenting an intriguing condensed-matter system to realize exotic many-body phenomena, as \textit{specifically} shown in “magic angle” twisted bilayer graphene for superconductivity and triangulene-based superatomic graphene for exciton condensation. Yet, no \textit{generic} physical model to realize such FBs has been developed. Here we present a new mathematical theorem, called bipartite double cover (BDC) theorem, and prove that the BDC of line-graph (LG) lattices hosts at least two chiral exact flat bands of opposite chirality, i.e., yin-yang FBs, centered-around/at charge neutrality ($E=0$) akin to the ``chiral limit" of twisted bilayer graphene. We illustrate this theorem by mapping it exactly onto tight-binding lattice models of the BDC of LGs of hexagonal lattice for strong topological and of triangular lattice for fragile topological FBs, respectively. Moreover, we use orbital design principle to realize such exotic yin-yang FBs in non-BDC lattices to instigate their real material discovery. This work not only enables the search for exact chiral FBs at zero energy beyond moiré heterostructures, but also opens the door to discovering quantum semiconductors featured with FB-enabled strongly correlated carriers.
\end{abstract}

\maketitle


In recent years, the importance of topological flat bands (FBs) in realizing exotic many-body phenomena, such as superconductivity \cite{1,2,3,4}, excitonic superfluidity \cite{5,6,7,8}, and magnetism \cite{2,9,10,11}, has been highlighted through a plethora of studies \cite{1,2,3,4,5,9,10,11,12,13,14,15} following “magic angle” twisted bilayer graphene (MATBG) \cite{1,16,17,18,19,20,21}. Lately, there has also been increasing interest in realizing chiral limit of MATBG with topologically fragile \textit{exact} FBs of opposite chirality at charge neutrality \cite{29,30,31,32} using external perturbations, such as periodic strain \cite{30}. Another intriguing case where topological FBs emerge near charge neutrality is a superatomic graphene lattice \cite{5,33,34,35}. It hosts two topological strong exact FBs of opposite chirality at $E=\pm t$, i.e., the yin-yang FBs centered symmetrically at $E=0$. These chiral exact FBs around/at charge neutrality exhibit fascinating transport properties of one/two-body carriers due to their overlapping Wannier functions, as manifested in unconventional superconductivity for fragile topological FBs of TBG \cite{13,36,37} and in excitonic Bose-Einstein condensation for strong topological yin-yang FBs of superatomic graphene \cite{5,6} respectively. \par
Beyond these specific systems, however, no generic theory or lattice design principle for constructing exact FBs with opposite chirality around/at Fermi level has been reported yet. On the other hand, topological FBs are also known to exist in a special class of two-dimensional (2D) lattices called line-graph (LG) lattices \cite{22,23,24,25,26}, where the presence of at least one FB is guaranteed by mathematical LG theorem \cite{27}. Compared to FBs at (around) $E=0$ in MATBG (superatomic graphene), however, FBs in LG lattices are at $E=-2t$ \cite{28} with $t$ being the tight-binding (TB) hopping integral. Although it was shown recently that LGs of so-called split-graph lattices contain one additional FB at $E=0$ \cite{26}, real material with split-graph lattice has been rarely found due to physically unrealistic hopping constraints.\par
In this article, we present a new graph theorem, called bipartite double cover (BDC) theorem and prove that the BDC of LG lattices always has at least two FBs of opposite chirality around/at the Fermi level formed by bonding and anti-bonding pairing of the two copies of LG FB eigenfunctions, respectively. We illustrate this theorem by mapping it exactly onto realistic physical models of TB Hamiltonian. First, we use the BDC of Kagome lattice (the LG of hexagonal lattice) to show that it has two strong topological FBs of opposite chirality as the valence and conduction band edge respectively forming a quantum semiconductor, as exemplified in the case of superatomic graphene \cite{6}. We then show that the BDC of LG of triangular lattice has four fragile topological exact FBs at charge neutrality, as exemplified in the case of chiral limit of MATBG \cite{29} but without the need of a superlattice potential or external fields. The FB topology of both BDC-LG lattices is confirmed using Chern number calculations and Wilson loop analyses.  Finally, we elaborate on effective orbital design principle \cite{23} for chiral exact FBs, based on the symmetries of BDC-LG wavefunctions to find orbitals that can be placed in the non-LG lattices to overcome the hopping constraints in the BDC-LG lattices, and hence to facilitate their material discovery.\par

\begin{figure}%
\centering
\includegraphics[width=0.48\textwidth]{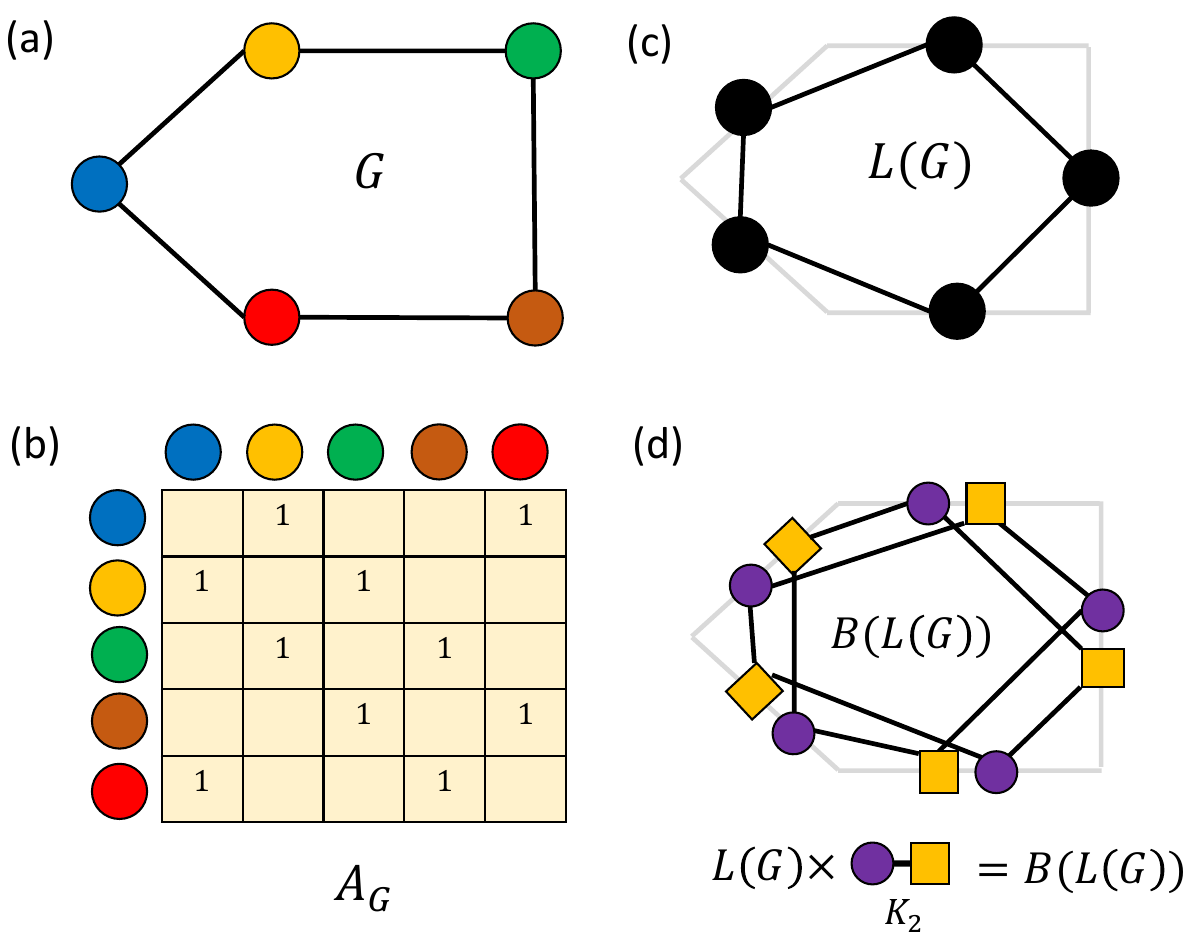}
\caption{(a) A finite cyclic graph ($G$) with 5 vertices indicated by colored circles and 5 edges indicated by black solid lines. (b) Adjacency matrix of $G$ ($A_G$) with matrix element 1 between the two circles which have an edge between them. (c) Line-graph of $G$ ($L(G)$). (d) BDC of $L(G)$ ($B(L(G))$).}\label{fig1}
\end{figure}
In mathematics, a graph ($G$) consists of a set of objects with some pairs of objects coupled with each other \cite{38}, as represented by vertices (objects) and edges (pair relation) in Fig.~\ref{fig1}(a). Each graph is assigned with an adjacency matrix ($A_G$), that has a matrix element of 1 to denote an adjacent pair of vertices (Fig.~\ref{fig1}(b)). An LG of $G$ ($L(G)$) is constructed to depict adjacencies between the edges of $G$. Namely, $L(G)$ is a graph, in which each vertex of $L(G)$ represents an edge of $G$, and the two vertices of $L(G)$ are adjacent if and only if their corresponding edges share a common vertex in $G$ (Fig.~\ref{fig1}(c)). $G$ is sometimes referred to as the root graph of $L(G)$. The LG theorem states - \textit{the spectra of infinite LG ($L(X)$), i.e., the eigenvalues of $A_{L(X)}$ with $X$ being an infinite graph, always has at least one lowest eigenvalue at -2 with infinite multiplicity provided the number of vertices exceeds the number of edges in the repeating unit of X} \cite{27}, which works in any dimensions, including 1D lattice \cite{23} (also see section I. in supplementary material (SM) \cite{46,59,61,62,58,alexandradinata2014wilson,zhou2019weyl}). \par 
The striking implication of LG theorem is seen in the electronic structure of lattices based on LGs. TB model of a lattice ($L$) is defined as the set of atomic sites and hopping matrix elements ($t$) between the sites. Each lattice can be associated with an infinite graph: vertex set of $X$ consisting of exactly one vertex for each lattice point, and an edge between all the vertex pairs for which $t\neq 0$. This leads to a one-to-one correspondence between a TB lattice and a mathematical graph \cite{28},
\begin{equation}
    H_L=tA_X,\label{eqn1}
\end{equation}
where $H_L$ denotes the TB Hamiltonian matrix in real space. Then, naturally, if $L$ is based on a LG, there exists at least one exact FB in the band structure at energy $E=-2t$. The degeneracy of the FB is given by the difference between the number of independent vertices and edges in the unit-cell (SM Section I. \cite{46}), which can also be explained by the index theorem or sublattice imbalance between the root graph and LG \cite{lieb1989two,ezawa2020systematic}. The LG FBs can be strong or fragile topological depending on if the root graph $X$ is bipartite or non-bipartite respectively \cite{26,39}. For the former, there is always a touching point between the FB and another dispersive band which can be viewed as a Berry flux center in analogy to Dirac point \cite{40} and be gapped out in the presence of spin-orbit coupling (SOC) leading to a strong topological quasi-FB, while for the latter, fragile exact FBs are isolated from rest of the bands \cite{39}, also at $E=-2t$. \par
In graph theory, a bipartite double cover (BDC) can also be constructed from a graph. BDC is one of the covering spaces of a graph, which is known to link discrete graphs to topological crystal structures \cite{41}. BDC maps every vertex $v_i$ of $G$ to two (double) vertices $u_i$ and $w_i$. It is called bipartite because the two vertices $u_i$ and $w_j$ are connected by an edge in BDC if and only if the corresponding $v_i$ and $v_j$ have an edge between them in $G$, i.e., one can divide the vertices of BDC into two independent subsets \cite{42}. In Fig.~\ref{fig1}(d), as an example, we show the BDC of a finite LG ($L(G)$) in Fig.~\ref{fig1}(b), denoted by $B(L(G))$. BDC can also be written as the tensor product of graphs, $L(G)\times K_2$, where $K_2$ is simply a two-vertex graph, consisting of a circle and a square in Fig.~\ref{fig1}(d). Edges exist only between a circle and a square and only if the two corresponding black circles have an edge between them in $L(G)$ (Fig.~\ref{fig1}(b)).\par
In this work we focus on the BDCs ($B(L(X))$) of infinite LGs ($L(X)$). Following the BDC construction, the adjacency matrix is given by,
\begin{equation}
    A_{B(L(X))}=\begin{bmatrix}
        0 & A_{L(X)}\\
        A_{L(X)}^T & 0\\
    \end{bmatrix},\label{eqn2}
\end{equation}
where $A_{L(X)}$ is the adjacency matrix of $L(X)$ and the basis of $A_{B(L(X))}$ is the set \{\{$u_i$\},\{$w_i$\}\} or \{\{circles\},\{squares\}\} so that the only non-zero matrix elements are in the off-diagonal blocks. Note that $A_{L(X)}$ is always a symmetric matrix, i.e., $A_{L(X)}=A_{L(X)}^T$ \cite{43}. Effectively, we have two copies of $L(X)$ but formed between two vertex subsets. Suppose one of the infinite eigenvectors of $A_{L(X)}$ corresponding to eigenvalue -2 is $v$ satisfying, $A_{L(X)}v=A_{L(X)}^Tv=-2v$. One can construct a vector $V=[v\;;\;v]$ in the basis of $A_{B(L(X))}$. From Eqn. (\ref{eqn2}), it is clear that $V$ will be an eigenvector of $A_{B(L(X)}$ satisfying $A_{B(L(X))}V=-2V$. Similarly, one can construct a vector $V'=[v\;;\;-v]$ that satisfies $A_{B(L(X))}V'=2V'$. Note that there are infinite possible $v$’s with eigenvalue -2 of $A_{L(X)}$, and for each of them one symmetrically ($V$)  and one anti-symmetrically ($V'$) paired eigenvectors of $A_{B(L(X))}$ can be constructed, formulating a \textit{new} graph theorem, which we call the “BDC theorem” – \textit{The spectra of bipartite double cover of infinite line-graphs contains at least two eigenvalues with infinite multiplicity at -2 and 2.}\par
One can physically understand the above theorem by carefully looking at the structure of $A_{B(L(X))}$. The absence of matrix element between the individual elements of {$u_i$} or {$w_i$} renders exact chiral symmetry to BDC which implies that if $\lambda$ is an eigenvalue of $A_{B(L(G))}$, its opposite $-\lambda$ should also be an eigenvalue. Alternatively, note that due to the chiral nature of $A_{B(L(X))}$, its square matrix $A_{B(L(X))}^2$ would be block diagonal given by,
\begin{equation}
    A_{B(L(X))}^2=\begin{bmatrix}
     A_{L(X)}^2 & 0 \\
     0 & A_{L(X)}^2
    \end{bmatrix}\label{eqn_sq}
\end{equation}
Now, if $\lambda$ is an eigenvalue of $A_{L(X)}$, $\lambda^2$ is an eigenvalue of $A_{L(X)}^2$. From Eqn. (\ref{eqn_sq}), it can be easily seen that $\lambda^2$ will also be an eigenvalue of $A_{B(L(X))}^2$. Then, using the property of square-root matrix \cite{ezawa2020systematic}, the eigenvalue set of $A_{B(L(X))}$ should consist of $\{\pm\lambda\}$, implying that the spectra of  $A_{B(L(X))}$ has two copies of the spectra of $A_{L(X)}$ with one inverted with respect to the other. Also, the eigenvectors corresponding to the two opposite eigenvalues are formed by the bonding ([$v$ ; $v$]) and anti-bonding ([$v$ ;$-v$]) pairing of the individual $L(X)$ eigenvectors. This theorem has important consequences on the electronic structure of a TB lattice associated with the BDC of a LG. It guarantees a chiral symmetric band structure having at least two FBs: one at $E=-2t$ and the other at $E=2t$.\par
\begin{figure}%
\centering
\includegraphics[width=0.46\textwidth]{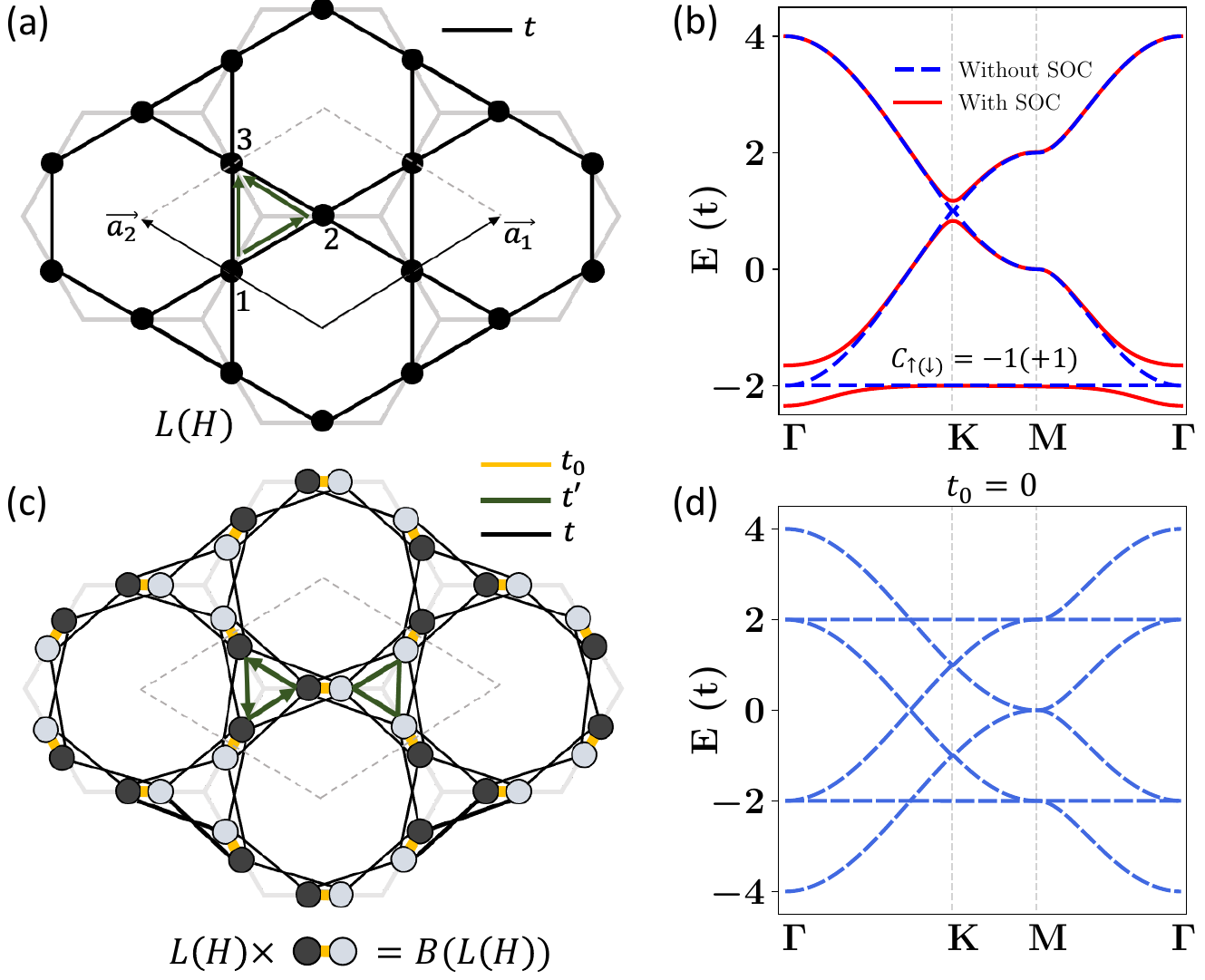}
\caption{(a) Line-graph of hexagonal lattice ($L(H)$) where gray lines form the root graph ($H$) while black lines form the Kagome lattice ($L(H)$). Unit-cell with three sites numbered 1-3 is depicted by gray dashed lines. Green arrows denote SOC hopping directions. (b) Band structure of (a) without (dashed blue lines) and with (solid red lines) SOC. (c) BDC construction of (a). Unit-cell is depicted by gray dashed lines. Green arrows are the directions of SOC hopping. (d) Band structure of (c) with $t_0=0$.}\label{fig2}
\end{figure}
To gain intuition about BDC construction and confirm the existence of two FBs in band structure, we revisit a well-known TB lattice model based on the LG of bipartite hexagonal lattice ($H$), i.e., Kagome lattice ($L(H)$), shown in Fig.~\ref{fig2}(a). The energy eigen spectrum of $H_{L(H)}$ is the typical Kagome band structure (dashed lines in Fig.~\ref{fig2}(b)), which has an exact FB at $E=-2t$ without SOC, in accordance with the LG theorem. Considering a spinful TB model with SOC ($\lambda$), a gap opens at the band touching point ($\Gamma$) leading to a topologically non-trivial quasi-FB (solid lines in Fig.~\ref{fig2}(b) at $\lambda=0.1t$) with Chern number $C_{\uparrow(\downarrow)}=-1(+1)$. The stable/strong topology of this FB can be confirmed by ribbon calculations where there are clear gapless edge states (Fig. S2(a) in the SM \cite{46}). \par
The BDC of Kagome lattice ($B(L(H))$) is constructed using the tensor product of $L(H)$ with a two-vertex graph ($K_2$) with dark- and light-filled circles in Fig.~\ref{fig2}(c). To maintain the translation symmetry, $K_2$ is placed along specific directions. Note that the first nearest neighbor (NN) hopping lines in $L(H)$ are transformed into third NN hopping lines in $B(L(H))$. The TB Hamiltonian can be directly constructed using Eqn. (\ref{eqn1}), and (\ref{eqn2}), leading to chiral symmetric electronic band structure as shown in Fig.~\ref{fig2}(d) having two exact FBs at $E=-2t$ and $E=2t$. However, the two FBs are not near/at charge neutrality as seen in the chiral limit of TBG or diatomic Kagome lattice \cite{33}. This can be remedied by noting that the hopping between dark and light filled atomic site in $B(L(H))$ is neglected. This is because the BDC construction does not have an edge between the vertices of $K_2$ (Fig.~\ref{fig1}(d)). Physically, this is the shortest bond length in a lattice and usually has a hopping integral ($t_0$, yellow bonds in Fig.~\ref{fig2}(c)) larger than the third NN hopping ($t$). An important observation here is that including $t_0$, i.e., including an edge between the two vertices of $K_2$, will not alter the chiral symmetry of $A_{B(L(H))}$ since this new edge introduces only a diagonal matrix element in the off-diagonal block of $A_{B(L(H))}$. Hence, the spinless TB Hamiltonian of $B(L(H))$ with the inclusion of $t_0$ can be written as,
\begin{equation}
    H_{B'(L(H))}=t A_{B(L(H))}+t_0 \begin{bmatrix}
        0 & I\\
        I & 0
    \end{bmatrix}.\label{eqn5}
\end{equation}
Where $I$ is identity matrix with the dimensions of the basis set of $L(H)$. The second term in Eqn.~\ref{eqn5} implies that the electronic spectrum of root $L(H)$ and of its chiral copy is shifted up- and down-wards by $t_0$, respectively, i.e., the band structure of $B'(L(H))$ has two FBs, but now at $E=-2t+t_0$, and $E=2t-t_0$.\par
\begin{figure}%
\centering
\includegraphics[width=0.48\textwidth]{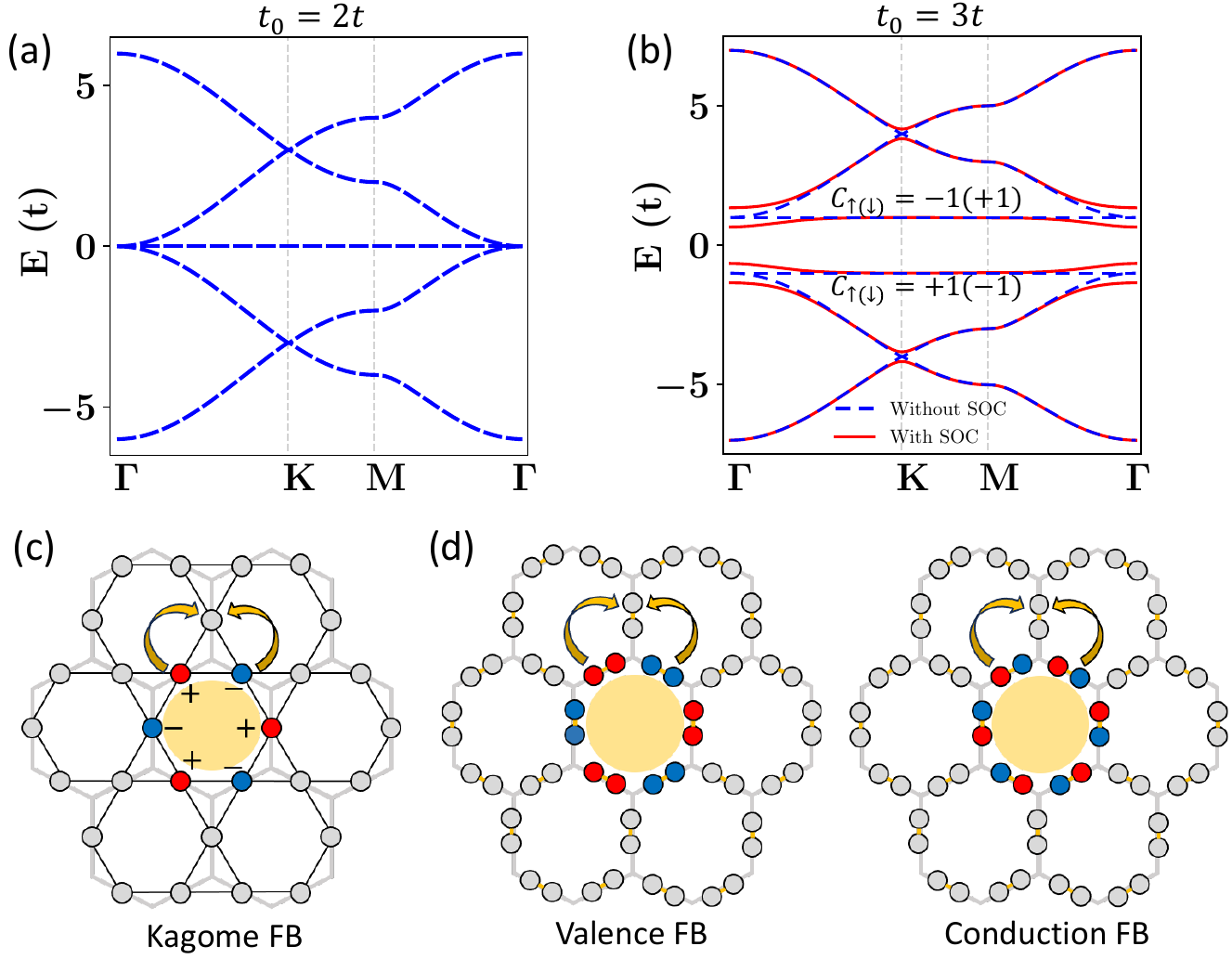}
\caption{(a) Band structure of Fig.~\ref{fig2}(c) with $t_0=2t$ containing two exact FBs at $E=0$. (b) Band structure of Fig.~\ref{fig2}(c) with $t_0=3t$. (c) Compact localized state (CLS) for Kagome FB in Fig.~\ref{fig2}(b). (d) CLS for yin-yang valence (left) and conduction (right) FB in (b) formed by bonding and anti-bonding pairing of (c) respectively. Blue and red circles indicate the opposite phases of the wavefunction respectively.}\label{fig3}
\end{figure}
In Fig.~\ref{fig3}(a) and \ref{fig3}(b) (dashed lines), we show the band structure obtained by diagonalizing $H_{B'(L(H))}$ with $t_0=2t$ and $t_0=3t$ respectively. For $t_0=2t$, there are two exact FBs completely degenerate at charge neutrality ($E=0$), while for $t_0=3t$, the two FBs are gapped ($E=\pm t$) forming valence and conduction band edges. Thus, the chiral symmetric band structures are guaranteed by the BDC theorem even with the inclusion of $t_0$. Importantly, both FBs have a band touching point with another dispersive band at $\Gamma$. Similar to the case of $L(H)$, they can be gapped out via SOC. A spinful Hamiltonian can be written by adding SOC effect containing positive (negative) hopping terms along (opposite to) the directions indicated by green dashed arrows in Fig.~\ref{fig2}(c). As seen from the solid lines in Fig.~\ref{fig3}(b), two quasi-FBs emerge with SOC having opposite chirality given by $C_{\uparrow(\downarrow)}^{FB_c}=-C_{\uparrow(\downarrow)}^{FB_v}=-1(+1)$, where $FB_{v(c)}$ indicates valence (conduction) FB. They are hence called yin-yang FBs, which have been recently shown to be an ideal platform for exotic optoelectronic properties, such as excitonic condensation \cite{6}, and giant circular dichroism \cite{44}. We have confirmed the strong topology of these FBs using ribbon calculations (Fig. S2(b) in SM \cite{46}).\par
An interesting feature of strong topological FBs is their compact localized (plaquette) states \cite{22}, manifesting a completely localized FB wavefunction and real-space topology, as shown in Fig.~\ref{fig3}(c) for Kagome FB. Outward hopping from the plaquette vanishes due to opposite phases of the wavefunction on neighboring sites that cancel each other out, leading to a destructive interference pattern. Interestingly, each of yin-yang FBs shows also the destructive interference patterns around a plaquette, but importantly with bonding and anti-bonding nature for the valence and conduction FB wavefunction respectively \cite{33}, as shown in Fig.~\ref{fig3}(d). Here we show that this is a direct consequence of BDC theorem involving bonding and anti-binding paring of the LG eigenfunctions.\par
\begin{figure}%
\centering
\includegraphics[width=0.48\textwidth]{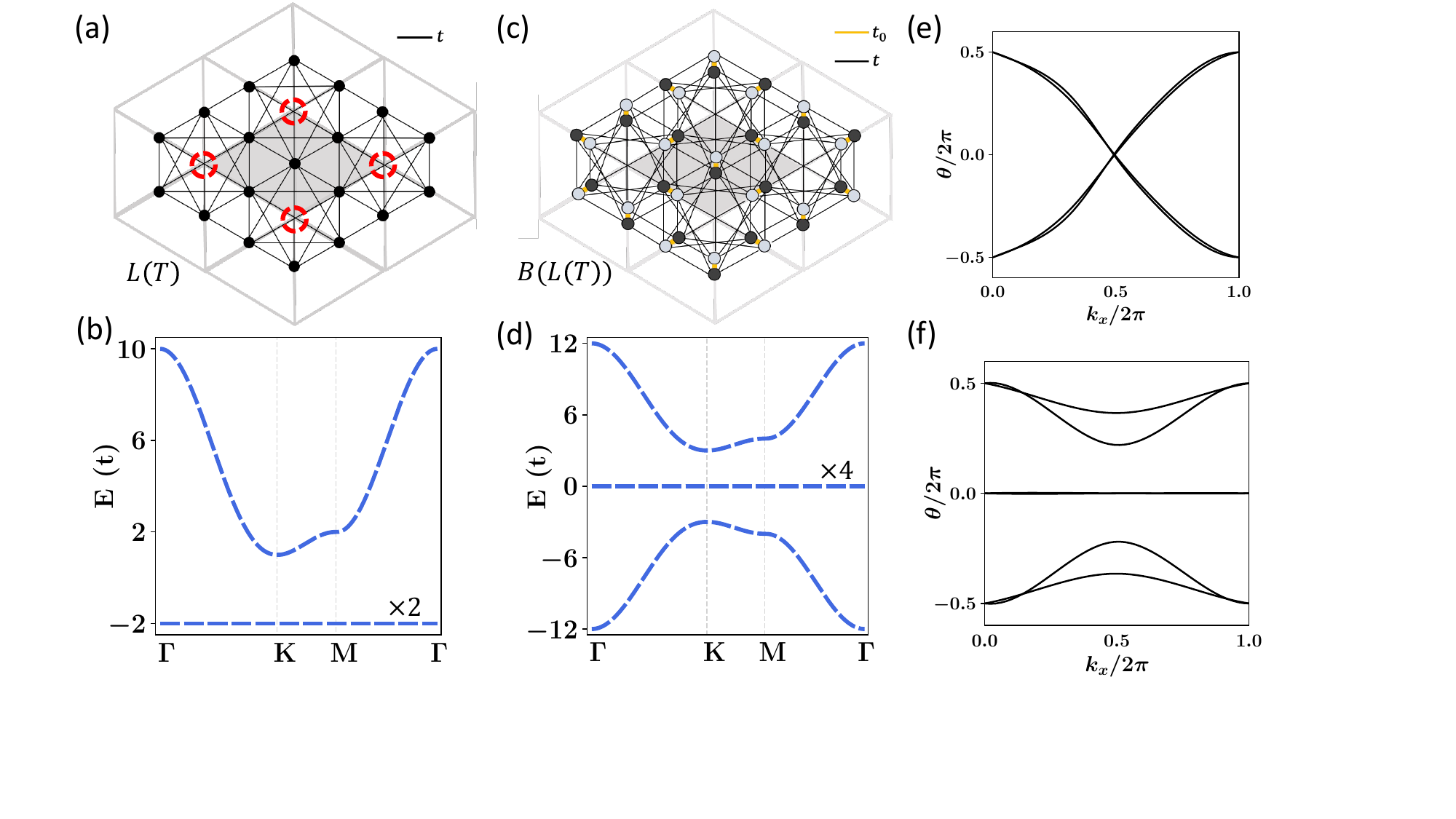}
\caption{(a) Kagome-3 lattice - Line-graph of triangular lattice ($L(T)$). Dashed red circle denotes a Wycoff position. (b) Band structure of (a). (c) BDC of (a) ($B(L(T))$). (d) Band structure of (c) with $t_0=2t$. (e). Wilson loop spectrum of the four FBs at charge neutrality in (d). (f) Same as (e) but with two additional $s$-orbitals in $B(L(T)$) at sites corresponding to the Wycoff position in (a).}\label{fig4}
\end{figure}
Next, we show the BDC construction for fragile topological FBs at charge neutrality. We consider the LG of non-bipartite triangular ($T$) lattice, so-called Kagome-3 lattice ($L(T)$) \cite{39,45} as depicted in Fig.~\ref{fig4}(a). Similar to $L(H)$, the unit-cell of $L(T)$ also contains three sites. The difference arises in the hopping terms. In $L(T)$, in addition to NN hopping, there are second and part of third NN hopping terms (Fig.~\ref{fig4}(a)) \cite{46}, leading to a TB band structure with two exact FBs at $E=-2t$ shown in Fig.~\ref{fig4}(b). The two FBs are fragile topological, i.e., topological obstruction to Wannier localization can be resolved by adding a trivial band \cite{47,48}. Previous studies \cite{39,45} have confirmed the fragile topology of these FBs using Wilson loop analysis \cite{49} which we include in SM \cite{46}. An essential point to note here is that the topology of these FBs at $E=-2t$ is akin to the single-particle bands of MATBG centered/at $E=0$ \cite{45}. Following the procedure illustrated above for strong topological FBs, one can also construct a BDC lattice of $L(T)$ as shown in Fig.~\ref{fig4}(c) in order to obtain fragile LG FBs centered/at $E=0$. The BDC construction leads to doubled sites in the unit-cell with a six-band TB Hamiltonian formulated following Eqn. (\ref{eqn1}) and (\ref{eqn2}) and the inclusion of $t_0$ (Eqn. (\ref{eqn5})) \cite{46}. The band-structure of $B(L(T))$ is shown in Fig.~\ref{fig4}(d) for $t_0=2t$. There are four exact FBs at charge neutrality, which are fragile topological. To illustrate this, in Fig.~\ref{fig4}(e) we calculate the Wilson loop of each spin sector separately, since the spinful time-reversal and $S^z$ symmetries are conserved. The winding in the Wilson loop spectra of the four FBs establishes their topological nature, without SOC, with two pairs each belonging to non-trivial Euler class $|e_2|=1$ \cite{50}. We next add trivial bands to the system by introducing an additional $s$-orbital on Wycoff position indicated by red circle in Fig.~\ref{fig4}(a) \cite{45}. The BDC of this extended model becomes an eight-band model \cite{46}. The Wilson loop spectra of six middle bands is shown in Fig.~\ref{fig4}(f). The non-trivial winding of FBs is now clearly removed confirming the fragile topology of these FBs at charge neutrality.\par
\begin{figure}%
\centering
\includegraphics[width=0.45\textwidth]{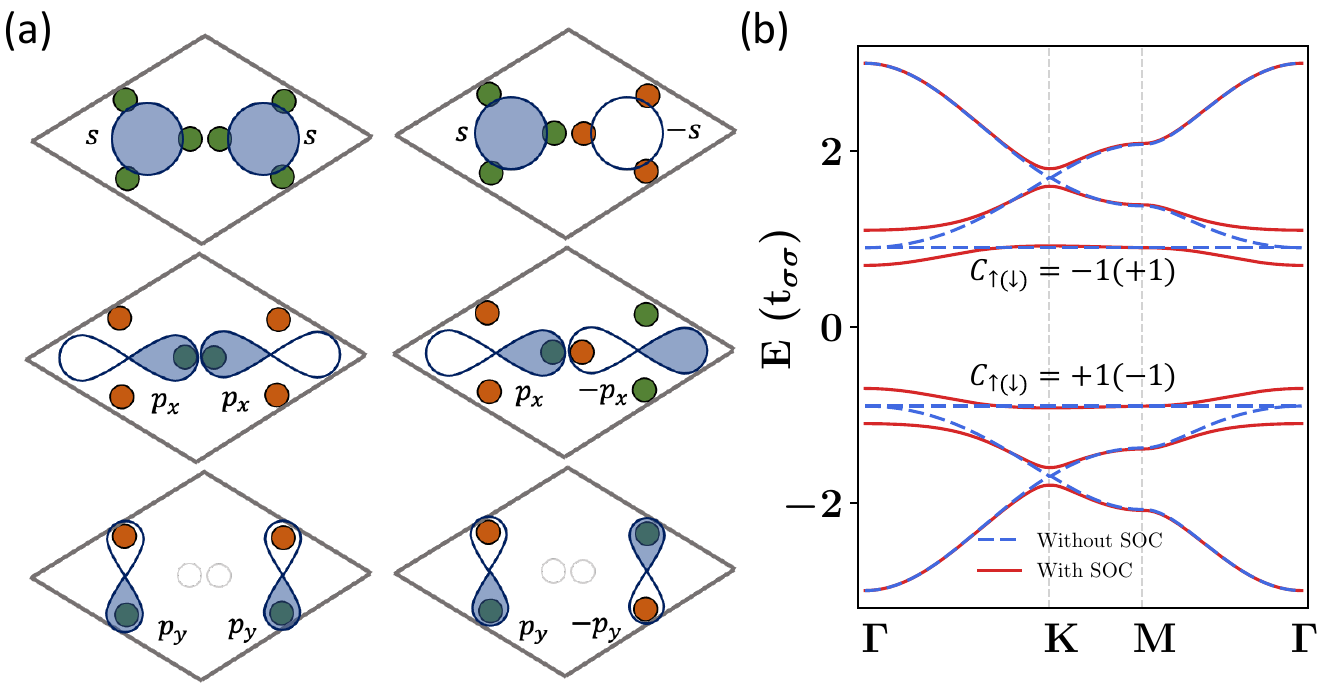}
\caption{(a) Six $\Gamma$ point wavefunctions of bands denoted by blue dashed lines in Fig.~\ref{fig3}(b). (b) TB band structure of hexagonal lattice model with $sp^2$ orbitals.}\label{fig5}
\end{figure}

Next, we notice that the BDC-LG lattice construction involves unusual hopping constraints, such as the absence of second NN hopping term ($t'=0$) in $B(L(H))$ (Fig.~\ref{fig2}(c)) while requiring a non-zero third NN hopping integral ($t\neq 0$). Realizing such peculiar hopping integrals in real materials could prove significantly challenging. On the other hand, most real materials are composed of multi atomic orbitals on each site instead of single $s$-orbital. Therefore, we finally illustrate orbital realization of chiral exact FBs of $B(L(H))$ following the procedure described in ref. \cite{23} that may instigate real material discovery of these exotic FBs. We first calculate the six eigenstates at $\Gamma$ of the $B(L(H))$ band structure (Fig.~\ref{fig3}(b)) and plot them in Fig.~\ref{fig5}(a). If one groups three sites into a “superatom” site, then the six sites of $B(L(H))$ can be mapped onto two sites of a hexagonal lattice with three orbitals on each site. And based on the symmetry of the eigenstates, six eigenstates in Fig.~\ref{fig5}(a) can be viewed as ($s, p_x, p_x$) orbitals on two hexagonal sites. Then, by choosing the two-center Slater-Koster hopping integrals \cite{51} as: $t_{ss\sigma}=1 eV$, $t_{sp\sigma}=0.8t_{ss\sigma}$, $t_{pp\sigma}=0.4t_{ss\sigma}$, in a spinless TB model \cite{46}, yin-yang FBs are reproduced as shown by dashed blue lines in Fig.~\ref{fig5}(b). We further add the onsite SOC on $p$ orbitals and consider a spinful model which opens a gap between dispersive band and FB (red solid lines in Fig.~\ref{fig5}(b). The calculated Chern numbers for two FBs are $C_{\uparrow(\downarrow)}^{FB_c}=-C_{\uparrow(\downarrow)}^{FB_v}=-1(+1)$. Note that the condition $t'=0$ in single $s$-orbital model transforms into the condition $t_{pp\pi}=0$ for $sp^2$-hexagonal model. Physically, $t_{pp\pi}$ is the weakest hopping integral by symmetry. A candidate real material should be one with $sp^2$ frontier orbitals on hexagonal symmetric lattice but a large lattice parameter which reduces $t_{pp\pi}$. Indeed, triangulene-based superatomic graphene satisfies this condition \cite{33} and has also been recently synthesized \cite{52}. Similar orbital design principle can be used for other BDC-LG constructions \cite{46}.\par
The BDC theorem, as derived and illustrated in this work, allows for the discovery of novel exact FBs at charge neutrality leading to an intrinsic lattice/material ($B(L(T)$) that can realize strongly correlated physics of MATBG without the need of external fields or fine-tuning of twist angles. Moreover, we point to the possibility of an unusual quantum semiconductor in $B(L(H))$ with valence and conduction FBs (yin-yang FBs), which upon doping create FB electron and hole carriers with unique transport properties due to their non-trivial quantum metric and non-zero superfluidic weight \cite{36}. This can have significant implications for the discovery and design of the beyond-Silicon electronic and optoelectronic devices. We emphasize that our BDC design principle works for all LG lattices, both in 2D and 3D \cite{46}. In addition, it can be easily extended to precursor-based organic materials such as triangulene crystals \cite{53}, and artificial lattice systems \cite{54,55,56}, where the hopping parameters can be manipulated by design. One can also consider weighted graphs with varying nearest neighbor hopping integrals \cite{28} to realize materials/models with flat bands of opposite chirality. Another possible future direction could be to investigate into the properties of the FBs in BDC lattices under external magnetic field \cite{aoki1996hofstadter,marques2023kaleidoscopes}.\par
$^{\dagger}$G.S., and B.X. contributed equally to this work. We acknowledge support from US Department of Energy (DOE)-Basic Energy Sciences (Grant No. DE-FG02- 04ER46148). All calculations were done on the CHPC at the University of Utah and NERSC.

\bibliography{manuscript}

\begin{thebibliography}{66}%
\makeatletter
\providecommand \@ifxundefined [1]{%
 \@ifx{#1\undefined}
}%
\providecommand \@ifnum [1]{%
 \ifnum #1\expandafter \@firstoftwo
 \else \expandafter \@secondoftwo
 \fi
}%
\providecommand \@ifx [1]{%
 \ifx #1\expandafter \@firstoftwo
 \else \expandafter \@secondoftwo
 \fi
}%
\providecommand \natexlab [1]{#1}%
\providecommand \enquote  [1]{``#1''}%
\providecommand \bibnamefont  [1]{#1}%
\providecommand \bibfnamefont [1]{#1}%
\providecommand \citenamefont [1]{#1}%
\providecommand \href@noop [0]{\@secondoftwo}%
\providecommand \href [0]{\begingroup \@sanitize@url \@href}%
\providecommand \@href[1]{\@@startlink{#1}\@@href}%
\providecommand \@@href[1]{\endgroup#1\@@endlink}%
\providecommand \@sanitize@url [0]{\catcode `\\12\catcode `\$12\catcode `\&12\catcode `\#12\catcode `\^12\catcode `\_12\catcode `\%12\relax}%
\providecommand \@@startlink[1]{}%
\providecommand \@@endlink[0]{}%
\providecommand \url  [0]{\begingroup\@sanitize@url \@url }%
\providecommand \@url [1]{\endgroup\@href {#1}{\urlprefix }}%
\providecommand \urlprefix  [0]{URL }%
\providecommand \Eprint [0]{\href }%
\providecommand \doibase [0]{https://doi.org/}%
\providecommand \selectlanguage [0]{\@gobble}%
\providecommand \bibinfo  [0]{\@secondoftwo}%
\providecommand \bibfield  [0]{\@secondoftwo}%
\providecommand \translation [1]{[#1]}%
\providecommand \BibitemOpen [0]{}%
\providecommand \bibitemStop [0]{}%
\providecommand \bibitemNoStop [0]{.\EOS\space}%
\providecommand \EOS [0]{\spacefactor3000\relax}%
\providecommand \BibitemShut  [1]{\csname bibitem#1\endcsname}%
\let\auto@bib@innerbib\@empty
\bibitem [{\citenamefont {Cao}\ \emph {et~al.}(2018)\citenamefont {Cao}, \citenamefont {Fatemi}, \citenamefont {Fang}, \citenamefont {Watanabe}, \citenamefont {Taniguchi}, \citenamefont {Kaxiras},\ and\ \citenamefont {Jarillo-Herrero}}]{1}%
  \BibitemOpen
  \bibfield  {author} {\bibinfo {author} {\bibfnamefont {Y.}~\bibnamefont {Cao}}, \bibinfo {author} {\bibfnamefont {V.}~\bibnamefont {Fatemi}}, \bibinfo {author} {\bibfnamefont {S.}~\bibnamefont {Fang}}, \bibinfo {author} {\bibfnamefont {K.}~\bibnamefont {Watanabe}}, \bibinfo {author} {\bibfnamefont {T.}~\bibnamefont {Taniguchi}}, \bibinfo {author} {\bibfnamefont {E.}~\bibnamefont {Kaxiras}},\ and\ \bibinfo {author} {\bibfnamefont {P.}~\bibnamefont {Jarillo-Herrero}},\ }\bibfield  {title} {\bibinfo {title} {Unconventional superconductivity in magic-angle graphene superlattices},\ }\href@noop {} {\bibfield  {journal} {\bibinfo  {journal} {Nature}\ }\textbf {\bibinfo {volume} {556}},\ \bibinfo {pages} {43} (\bibinfo {year} {2018})}\BibitemShut {NoStop}%
\bibitem [{\citenamefont {Balents}\ \emph {et~al.}(2020)\citenamefont {Balents}, \citenamefont {Dean}, \citenamefont {Efetov},\ and\ \citenamefont {Young}}]{2}%
  \BibitemOpen
  \bibfield  {author} {\bibinfo {author} {\bibfnamefont {L.}~\bibnamefont {Balents}}, \bibinfo {author} {\bibfnamefont {C.~R.}\ \bibnamefont {Dean}}, \bibinfo {author} {\bibfnamefont {D.~K.}\ \bibnamefont {Efetov}},\ and\ \bibinfo {author} {\bibfnamefont {A.~F.}\ \bibnamefont {Young}},\ }\bibfield  {title} {\bibinfo {title} {Superconductivity and strong correlations in moir{\'e} flat bands},\ }\href@noop {} {\bibfield  {journal} {\bibinfo  {journal} {Nature Physics}\ }\textbf {\bibinfo {volume} {16}},\ \bibinfo {pages} {725} (\bibinfo {year} {2020})}\BibitemShut {NoStop}%
\bibitem [{\citenamefont {Park}\ \emph {et~al.}(2021)\citenamefont {Park}, \citenamefont {Cao}, \citenamefont {Watanabe}, \citenamefont {Taniguchi},\ and\ \citenamefont {Jarillo-Herrero}}]{3}%
  \BibitemOpen
  \bibfield  {author} {\bibinfo {author} {\bibfnamefont {J.~M.}\ \bibnamefont {Park}}, \bibinfo {author} {\bibfnamefont {Y.}~\bibnamefont {Cao}}, \bibinfo {author} {\bibfnamefont {K.}~\bibnamefont {Watanabe}}, \bibinfo {author} {\bibfnamefont {T.}~\bibnamefont {Taniguchi}},\ and\ \bibinfo {author} {\bibfnamefont {P.}~\bibnamefont {Jarillo-Herrero}},\ }\bibfield  {title} {\bibinfo {title} {Tunable strongly coupled superconductivity in magic-angle twisted trilayer graphene},\ }\href@noop {} {\bibfield  {journal} {\bibinfo  {journal} {Nature}\ }\textbf {\bibinfo {volume} {590}},\ \bibinfo {pages} {249} (\bibinfo {year} {2021})}\BibitemShut {NoStop}%
\bibitem [{\citenamefont {Park}\ \emph {et~al.}(2022)\citenamefont {Park}, \citenamefont {Cao}, \citenamefont {Xia}, \citenamefont {Sun}, \citenamefont {Watanabe}, \citenamefont {Taniguchi},\ and\ \citenamefont {Jarillo-Herrero}}]{4}%
  \BibitemOpen
  \bibfield  {author} {\bibinfo {author} {\bibfnamefont {J.~M.}\ \bibnamefont {Park}}, \bibinfo {author} {\bibfnamefont {Y.}~\bibnamefont {Cao}}, \bibinfo {author} {\bibfnamefont {L.-Q.}\ \bibnamefont {Xia}}, \bibinfo {author} {\bibfnamefont {S.}~\bibnamefont {Sun}}, \bibinfo {author} {\bibfnamefont {K.}~\bibnamefont {Watanabe}}, \bibinfo {author} {\bibfnamefont {T.}~\bibnamefont {Taniguchi}},\ and\ \bibinfo {author} {\bibfnamefont {P.}~\bibnamefont {Jarillo-Herrero}},\ }\bibfield  {title} {\bibinfo {title} {Robust superconductivity in magic-angle multilayer graphene family},\ }\href@noop {} {\bibfield  {journal} {\bibinfo  {journal} {Nature Materials}\ }\textbf {\bibinfo {volume} {21}},\ \bibinfo {pages} {877} (\bibinfo {year} {2022})}\BibitemShut {NoStop}%
\bibitem [{\citenamefont {Sethi}\ \emph {et~al.}(2021)\citenamefont {Sethi}, \citenamefont {Zhou}, \citenamefont {Zhu}, \citenamefont {Yang},\ and\ \citenamefont {Liu}}]{5}%
  \BibitemOpen
  \bibfield  {author} {\bibinfo {author} {\bibfnamefont {G.}~\bibnamefont {Sethi}}, \bibinfo {author} {\bibfnamefont {Y.}~\bibnamefont {Zhou}}, \bibinfo {author} {\bibfnamefont {L.}~\bibnamefont {Zhu}}, \bibinfo {author} {\bibfnamefont {L.}~\bibnamefont {Yang}},\ and\ \bibinfo {author} {\bibfnamefont {F.}~\bibnamefont {Liu}},\ }\bibfield  {title} {\bibinfo {title} {Flat-band-enabled triplet excitonic insulator in a diatomic kagome lattice},\ }\href@noop {} {\bibfield  {journal} {\bibinfo  {journal} {Physical Review Letters}\ }\textbf {\bibinfo {volume} {126}},\ \bibinfo {pages} {196403} (\bibinfo {year} {2021})}\BibitemShut {NoStop}%
\bibitem [{\citenamefont {Sethi}\ \emph {et~al.}(2023)\citenamefont {Sethi}, \citenamefont {Cuma},\ and\ \citenamefont {Liu}}]{6}%
  \BibitemOpen
  \bibfield  {author} {\bibinfo {author} {\bibfnamefont {G.}~\bibnamefont {Sethi}}, \bibinfo {author} {\bibfnamefont {M.}~\bibnamefont {Cuma}},\ and\ \bibinfo {author} {\bibfnamefont {F.}~\bibnamefont {Liu}},\ }\bibfield  {title} {\bibinfo {title} {Excitonic condensate in flat valence and conduction bands of opposite chirality},\ }\href@noop {} {\bibfield  {journal} {\bibinfo  {journal} {Physical Review Letters}\ }\textbf {\bibinfo {volume} {130}},\ \bibinfo {pages} {186401} (\bibinfo {year} {2023})}\BibitemShut {NoStop}%
\bibitem [{\citenamefont {Sethi}\ \emph {et~al.}(2022)\citenamefont {Sethi}, \citenamefont {Sheng},\ and\ \citenamefont {Liu}}]{7}%
  \BibitemOpen
  \bibfield  {author} {\bibinfo {author} {\bibfnamefont {G.}~\bibnamefont {Sethi}}, \bibinfo {author} {\bibfnamefont {D.}~\bibnamefont {Sheng}},\ and\ \bibinfo {author} {\bibfnamefont {F.}~\bibnamefont {Liu}},\ }\bibfield  {title} {\bibinfo {title} {Anomalous bilayer quantum hall effect},\ }\href@noop {} {\bibfield  {journal} {\bibinfo  {journal} {arXiv preprint arXiv:2211.04613}\ } (\bibinfo {year} {2022})}\BibitemShut {NoStop}%
\bibitem [{\citenamefont {Hu}\ \emph {et~al.}(2022)\citenamefont {Hu}, \citenamefont {Hyart}, \citenamefont {Pikulin},\ and\ \citenamefont {Rossi}}]{8}%
  \BibitemOpen
  \bibfield  {author} {\bibinfo {author} {\bibfnamefont {X.}~\bibnamefont {Hu}}, \bibinfo {author} {\bibfnamefont {T.}~\bibnamefont {Hyart}}, \bibinfo {author} {\bibfnamefont {D.~I.}\ \bibnamefont {Pikulin}},\ and\ \bibinfo {author} {\bibfnamefont {E.}~\bibnamefont {Rossi}},\ }\bibfield  {title} {\bibinfo {title} {Quantum-metric-enabled exciton condensate in double twisted bilayer graphene},\ }\href@noop {} {\bibfield  {journal} {\bibinfo  {journal} {Physical Review B}\ }\textbf {\bibinfo {volume} {105}},\ \bibinfo {pages} {L140506} (\bibinfo {year} {2022})}\BibitemShut {NoStop}%
\bibitem [{\citenamefont {Liu}\ and\ \citenamefont {Dai}(2021)}]{9}%
  \BibitemOpen
  \bibfield  {author} {\bibinfo {author} {\bibfnamefont {J.}~\bibnamefont {Liu}}\ and\ \bibinfo {author} {\bibfnamefont {X.}~\bibnamefont {Dai}},\ }\bibfield  {title} {\bibinfo {title} {Orbital magnetic states in moir{\'e} graphene systems},\ }\href@noop {} {\bibfield  {journal} {\bibinfo  {journal} {Nature Reviews Physics}\ }\textbf {\bibinfo {volume} {3}},\ \bibinfo {pages} {367} (\bibinfo {year} {2021})}\BibitemShut {NoStop}%
\bibitem [{\citenamefont {Tschirhart}\ \emph {et~al.}(2021)\citenamefont {Tschirhart}, \citenamefont {Serlin}, \citenamefont {Polshyn}, \citenamefont {Shragai}, \citenamefont {Xia}, \citenamefont {Zhu}, \citenamefont {Zhang}, \citenamefont {Watanabe}, \citenamefont {Taniguchi}, \citenamefont {Huber} \emph {et~al.}}]{10}%
  \BibitemOpen
  \bibfield  {author} {\bibinfo {author} {\bibfnamefont {C.}~\bibnamefont {Tschirhart}}, \bibinfo {author} {\bibfnamefont {M.}~\bibnamefont {Serlin}}, \bibinfo {author} {\bibfnamefont {H.}~\bibnamefont {Polshyn}}, \bibinfo {author} {\bibfnamefont {A.}~\bibnamefont {Shragai}}, \bibinfo {author} {\bibfnamefont {Z.}~\bibnamefont {Xia}}, \bibinfo {author} {\bibfnamefont {J.}~\bibnamefont {Zhu}}, \bibinfo {author} {\bibfnamefont {Y.}~\bibnamefont {Zhang}}, \bibinfo {author} {\bibfnamefont {K.}~\bibnamefont {Watanabe}}, \bibinfo {author} {\bibfnamefont {T.}~\bibnamefont {Taniguchi}}, \bibinfo {author} {\bibfnamefont {M.}~\bibnamefont {Huber}}, \emph {et~al.},\ }\bibfield  {title} {\bibinfo {title} {Imaging orbital ferromagnetism in a moir{\'e} chern insulator},\ }\href@noop {} {\bibfield  {journal} {\bibinfo  {journal} {Science}\ }\textbf {\bibinfo {volume} {372}},\ \bibinfo {pages} {1323} (\bibinfo {year} {2021})}\BibitemShut {NoStop}%
\bibitem [{\citenamefont {Andrei}\ \emph {et~al.}(2021)\citenamefont {Andrei}, \citenamefont {Efetov}, \citenamefont {Jarillo-Herrero}, \citenamefont {MacDonald}, \citenamefont {Mak}, \citenamefont {Senthil}, \citenamefont {Tutuc}, \citenamefont {Yazdani},\ and\ \citenamefont {Young}}]{11}%
  \BibitemOpen
  \bibfield  {author} {\bibinfo {author} {\bibfnamefont {E.~Y.}\ \bibnamefont {Andrei}}, \bibinfo {author} {\bibfnamefont {D.~K.}\ \bibnamefont {Efetov}}, \bibinfo {author} {\bibfnamefont {P.}~\bibnamefont {Jarillo-Herrero}}, \bibinfo {author} {\bibfnamefont {A.~H.}\ \bibnamefont {MacDonald}}, \bibinfo {author} {\bibfnamefont {K.~F.}\ \bibnamefont {Mak}}, \bibinfo {author} {\bibfnamefont {T.}~\bibnamefont {Senthil}}, \bibinfo {author} {\bibfnamefont {E.}~\bibnamefont {Tutuc}}, \bibinfo {author} {\bibfnamefont {A.}~\bibnamefont {Yazdani}},\ and\ \bibinfo {author} {\bibfnamefont {A.~F.}\ \bibnamefont {Young}},\ }\bibfield  {title} {\bibinfo {title} {The marvels of moir{\'e} materials},\ }\href@noop {} {\bibfield  {journal} {\bibinfo  {journal} {Nature Reviews Materials}\ }\textbf {\bibinfo {volume} {6}},\ \bibinfo {pages} {201} (\bibinfo {year} {2021})}\BibitemShut {NoStop}%
\bibitem [{\citenamefont {Zhang}\ \emph {et~al.}(2019)\citenamefont {Zhang}, \citenamefont {Mao}, \citenamefont {Cao}, \citenamefont {Jarillo-Herrero},\ and\ \citenamefont {Senthil}}]{12}%
  \BibitemOpen
  \bibfield  {author} {\bibinfo {author} {\bibfnamefont {Y.-H.}\ \bibnamefont {Zhang}}, \bibinfo {author} {\bibfnamefont {D.}~\bibnamefont {Mao}}, \bibinfo {author} {\bibfnamefont {Y.}~\bibnamefont {Cao}}, \bibinfo {author} {\bibfnamefont {P.}~\bibnamefont {Jarillo-Herrero}},\ and\ \bibinfo {author} {\bibfnamefont {T.}~\bibnamefont {Senthil}},\ }\bibfield  {title} {\bibinfo {title} {Nearly flat chern bands in moir{\'e} superlattices},\ }\href@noop {} {\bibfield  {journal} {\bibinfo  {journal} {Physical Review B}\ }\textbf {\bibinfo {volume} {99}},\ \bibinfo {pages} {075127} (\bibinfo {year} {2019})}\BibitemShut {NoStop}%
\bibitem [{\citenamefont {T{\"o}rm{\"a}}\ \emph {et~al.}(2018)\citenamefont {T{\"o}rm{\"a}}, \citenamefont {Liang},\ and\ \citenamefont {Peotta}}]{13}%
  \BibitemOpen
  \bibfield  {author} {\bibinfo {author} {\bibfnamefont {P.}~\bibnamefont {T{\"o}rm{\"a}}}, \bibinfo {author} {\bibfnamefont {L.}~\bibnamefont {Liang}},\ and\ \bibinfo {author} {\bibfnamefont {S.}~\bibnamefont {Peotta}},\ }\bibfield  {title} {\bibinfo {title} {Quantum metric and effective mass of a two-body bound state in a flat band},\ }\href@noop {} {\bibfield  {journal} {\bibinfo  {journal} {Physical Review B}\ }\textbf {\bibinfo {volume} {98}},\ \bibinfo {pages} {220511} (\bibinfo {year} {2018})}\BibitemShut {NoStop}%
\bibitem [{\citenamefont {Kwan}\ \emph {et~al.}(2021)\citenamefont {Kwan}, \citenamefont {Hu}, \citenamefont {Simon},\ and\ \citenamefont {Parameswaran}}]{14}%
  \BibitemOpen
  \bibfield  {author} {\bibinfo {author} {\bibfnamefont {Y.~H.}\ \bibnamefont {Kwan}}, \bibinfo {author} {\bibfnamefont {Y.}~\bibnamefont {Hu}}, \bibinfo {author} {\bibfnamefont {S.~H.}\ \bibnamefont {Simon}},\ and\ \bibinfo {author} {\bibfnamefont {S.}~\bibnamefont {Parameswaran}},\ }\bibfield  {title} {\bibinfo {title} {Exciton band topology in spontaneous quantum anomalous hall insulators: Applications to twisted bilayer graphene},\ }\href@noop {} {\bibfield  {journal} {\bibinfo  {journal} {Physical Review Letters}\ }\textbf {\bibinfo {volume} {126}},\ \bibinfo {pages} {137601} (\bibinfo {year} {2021})}\BibitemShut {NoStop}%
\bibitem [{\citenamefont {Song}\ and\ \citenamefont {Bernevig}(2022)}]{15}%
  \BibitemOpen
  \bibfield  {author} {\bibinfo {author} {\bibfnamefont {Z.-D.}\ \bibnamefont {Song}}\ and\ \bibinfo {author} {\bibfnamefont {B.~A.}\ \bibnamefont {Bernevig}},\ }\bibfield  {title} {\bibinfo {title} {Magic-angle twisted bilayer graphene as a topological heavy fermion problem},\ }\href@noop {} {\bibfield  {journal} {\bibinfo  {journal} {Physical review letters}\ }\textbf {\bibinfo {volume} {129}},\ \bibinfo {pages} {047601} (\bibinfo {year} {2022})}\BibitemShut {NoStop}%
\bibitem [{\citenamefont {Kennes}\ \emph {et~al.}(2021)\citenamefont {Kennes}, \citenamefont {Claassen}, \citenamefont {Xian}, \citenamefont {Georges}, \citenamefont {Millis}, \citenamefont {Hone}, \citenamefont {Dean}, \citenamefont {Basov}, \citenamefont {Pasupathy},\ and\ \citenamefont {Rubio}}]{16}%
  \BibitemOpen
  \bibfield  {author} {\bibinfo {author} {\bibfnamefont {D.~M.}\ \bibnamefont {Kennes}}, \bibinfo {author} {\bibfnamefont {M.}~\bibnamefont {Claassen}}, \bibinfo {author} {\bibfnamefont {L.}~\bibnamefont {Xian}}, \bibinfo {author} {\bibfnamefont {A.}~\bibnamefont {Georges}}, \bibinfo {author} {\bibfnamefont {A.~J.}\ \bibnamefont {Millis}}, \bibinfo {author} {\bibfnamefont {J.}~\bibnamefont {Hone}}, \bibinfo {author} {\bibfnamefont {C.~R.}\ \bibnamefont {Dean}}, \bibinfo {author} {\bibfnamefont {D.}~\bibnamefont {Basov}}, \bibinfo {author} {\bibfnamefont {A.~N.}\ \bibnamefont {Pasupathy}},\ and\ \bibinfo {author} {\bibfnamefont {A.}~\bibnamefont {Rubio}},\ }\bibfield  {title} {\bibinfo {title} {Moir{\'e} heterostructures as a condensed-matter quantum simulator},\ }\href@noop {} {\ \textbf {\bibinfo {volume} {17}},\ \bibinfo {pages} {155} (\bibinfo {year} {2021})}\BibitemShut {NoStop}%
\bibitem [{\citenamefont {Mak}\ and\ \citenamefont {Shan}(2022)}]{17}%
  \BibitemOpen
  \bibfield  {author} {\bibinfo {author} {\bibfnamefont {K.~F.}\ \bibnamefont {Mak}}\ and\ \bibinfo {author} {\bibfnamefont {J.}~\bibnamefont {Shan}},\ }\bibfield  {title} {\bibinfo {title} {Semiconductor moir{\'e} materials},\ }\href@noop {} {\bibfield  {journal} {\bibinfo  {journal} {Nature Nanotechnology}\ }\textbf {\bibinfo {volume} {17}},\ \bibinfo {pages} {686} (\bibinfo {year} {2022})}\BibitemShut {NoStop}%
\bibitem [{\citenamefont {Naik}\ \emph {et~al.}(2022)\citenamefont {Naik}, \citenamefont {Regan}, \citenamefont {Zhang}, \citenamefont {Chan}, \citenamefont {Li}, \citenamefont {Wang}, \citenamefont {Yoon}, \citenamefont {Ong}, \citenamefont {Zhao}, \citenamefont {Zhao} \emph {et~al.}}]{18}%
  \BibitemOpen
  \bibfield  {author} {\bibinfo {author} {\bibfnamefont {M.~H.}\ \bibnamefont {Naik}}, \bibinfo {author} {\bibfnamefont {E.~C.}\ \bibnamefont {Regan}}, \bibinfo {author} {\bibfnamefont {Z.}~\bibnamefont {Zhang}}, \bibinfo {author} {\bibfnamefont {Y.-H.}\ \bibnamefont {Chan}}, \bibinfo {author} {\bibfnamefont {Z.}~\bibnamefont {Li}}, \bibinfo {author} {\bibfnamefont {D.}~\bibnamefont {Wang}}, \bibinfo {author} {\bibfnamefont {Y.}~\bibnamefont {Yoon}}, \bibinfo {author} {\bibfnamefont {C.~S.}\ \bibnamefont {Ong}}, \bibinfo {author} {\bibfnamefont {W.}~\bibnamefont {Zhao}}, \bibinfo {author} {\bibfnamefont {S.}~\bibnamefont {Zhao}}, \emph {et~al.},\ }\bibfield  {title} {\bibinfo {title} {Intralayer charge-transfer moir{\'e} excitons in van der waals superlattices},\ }\href@noop {} {\bibfield  {journal} {\bibinfo  {journal} {Nature}\ }\textbf {\bibinfo {volume} {609}},\ \bibinfo {pages} {52} (\bibinfo {year} {2022})}\BibitemShut {NoStop}%
\bibitem [{\citenamefont {Wu}\ \emph {et~al.}(2018)\citenamefont {Wu}, \citenamefont {Lovorn}, \citenamefont {Tutuc},\ and\ \citenamefont {MacDonald}}]{19}%
  \BibitemOpen
  \bibfield  {author} {\bibinfo {author} {\bibfnamefont {F.}~\bibnamefont {Wu}}, \bibinfo {author} {\bibfnamefont {T.}~\bibnamefont {Lovorn}}, \bibinfo {author} {\bibfnamefont {E.}~\bibnamefont {Tutuc}},\ and\ \bibinfo {author} {\bibfnamefont {A.~H.}\ \bibnamefont {MacDonald}},\ }\bibfield  {title} {\bibinfo {title} {Hubbard model physics in transition metal dichalcogenide moir{\'e} bands},\ }\href@noop {} {\bibfield  {journal} {\bibinfo  {journal} {Physical review letters}\ }\textbf {\bibinfo {volume} {121}},\ \bibinfo {pages} {026402} (\bibinfo {year} {2018})}\BibitemShut {NoStop}%
\bibitem [{\citenamefont {Bistritzer}\ and\ \citenamefont {MacDonald}(2011)}]{20}%
  \BibitemOpen
  \bibfield  {author} {\bibinfo {author} {\bibfnamefont {R.}~\bibnamefont {Bistritzer}}\ and\ \bibinfo {author} {\bibfnamefont {A.~H.}\ \bibnamefont {MacDonald}},\ }\bibfield  {title} {\bibinfo {title} {Moir{\'e} bands in twisted double-layer graphene},\ }\href@noop {} {\bibfield  {journal} {\bibinfo  {journal} {Proceedings of the National Academy of Sciences}\ }\textbf {\bibinfo {volume} {108}},\ \bibinfo {pages} {12233} (\bibinfo {year} {2011})}\BibitemShut {NoStop}%
\bibitem [{\citenamefont {Dos~Santos}\ \emph {et~al.}(2012)\citenamefont {Dos~Santos}, \citenamefont {Peres},\ and\ \citenamefont {Neto}}]{21}%
  \BibitemOpen
  \bibfield  {author} {\bibinfo {author} {\bibfnamefont {J.~L.}\ \bibnamefont {Dos~Santos}}, \bibinfo {author} {\bibfnamefont {N.}~\bibnamefont {Peres}},\ and\ \bibinfo {author} {\bibfnamefont {A.~C.}\ \bibnamefont {Neto}},\ }\bibfield  {title} {\bibinfo {title} {Continuum model of the twisted graphene bilayer},\ }\href@noop {} {\bibfield  {journal} {\bibinfo  {journal} {Physical review B}\ }\textbf {\bibinfo {volume} {86}},\ \bibinfo {pages} {155449} (\bibinfo {year} {2012})}\BibitemShut {NoStop}%
\bibitem [{\citenamefont {Tarnopolsky}\ \emph {et~al.}(2019)\citenamefont {Tarnopolsky}, \citenamefont {Kruchkov},\ and\ \citenamefont {Vishwanath}}]{29}%
  \BibitemOpen
  \bibfield  {author} {\bibinfo {author} {\bibfnamefont {G.}~\bibnamefont {Tarnopolsky}}, \bibinfo {author} {\bibfnamefont {A.~J.}\ \bibnamefont {Kruchkov}},\ and\ \bibinfo {author} {\bibfnamefont {A.}~\bibnamefont {Vishwanath}},\ }\bibfield  {title} {\bibinfo {title} {Origin of magic angles in twisted bilayer graphene},\ }\href@noop {} {\bibfield  {journal} {\bibinfo  {journal} {Physical review letters}\ }\textbf {\bibinfo {volume} {122}},\ \bibinfo {pages} {106405} (\bibinfo {year} {2019})}\BibitemShut {NoStop}%
\bibitem [{\citenamefont {Wan}\ \emph {et~al.}(2023)\citenamefont {Wan}, \citenamefont {Sarkar}, \citenamefont {Lin},\ and\ \citenamefont {Sun}}]{30}%
  \BibitemOpen
  \bibfield  {author} {\bibinfo {author} {\bibfnamefont {X.}~\bibnamefont {Wan}}, \bibinfo {author} {\bibfnamefont {S.}~\bibnamefont {Sarkar}}, \bibinfo {author} {\bibfnamefont {S.-Z.}\ \bibnamefont {Lin}},\ and\ \bibinfo {author} {\bibfnamefont {K.}~\bibnamefont {Sun}},\ }\bibfield  {title} {\bibinfo {title} {Topological exact flat bands in two-dimensional materials under periodic strain},\ }\href@noop {} {\bibfield  {journal} {\bibinfo  {journal} {Physical Review Letters}\ }\textbf {\bibinfo {volume} {130}},\ \bibinfo {pages} {216401} (\bibinfo {year} {2023})}\BibitemShut {NoStop}%
\bibitem [{\citenamefont {Li}\ \emph {et~al.}(2022)\citenamefont {Li}, \citenamefont {He},\ and\ \citenamefont {Yao}}]{31}%
  \BibitemOpen
  \bibfield  {author} {\bibinfo {author} {\bibfnamefont {M.-R.}\ \bibnamefont {Li}}, \bibinfo {author} {\bibfnamefont {A.-L.}\ \bibnamefont {He}},\ and\ \bibinfo {author} {\bibfnamefont {H.}~\bibnamefont {Yao}},\ }\bibfield  {title} {\bibinfo {title} {Magic-angle twisted bilayer systems with quadratic band touching: Exactly flat bands with high chern number},\ }\href@noop {} {\bibfield  {journal} {\bibinfo  {journal} {Physical Review Research}\ }\textbf {\bibinfo {volume} {4}},\ \bibinfo {pages} {043151} (\bibinfo {year} {2022})}\BibitemShut {NoStop}%
\bibitem [{\citenamefont {Gao}\ \emph {et~al.}(2022)\citenamefont {Gao}, \citenamefont {Dong}, \citenamefont {Ledwith}, \citenamefont {Parker},\ and\ \citenamefont {Khalaf}}]{32}%
  \BibitemOpen
  \bibfield  {author} {\bibinfo {author} {\bibfnamefont {Q.}~\bibnamefont {Gao}}, \bibinfo {author} {\bibfnamefont {J.}~\bibnamefont {Dong}}, \bibinfo {author} {\bibfnamefont {P.}~\bibnamefont {Ledwith}}, \bibinfo {author} {\bibfnamefont {D.}~\bibnamefont {Parker}},\ and\ \bibinfo {author} {\bibfnamefont {E.}~\bibnamefont {Khalaf}},\ }\bibfield  {title} {\bibinfo {title} {Untwisting moir{\'e} physics: Almost ideal bands and fractional chern insulators in periodically strained monolayer graphene},\ }\href@noop {} {\bibfield  {journal} {\bibinfo  {journal} {arXiv e-prints}\ ,\ \bibinfo {pages} {arXiv}} (\bibinfo {year} {2022})}\BibitemShut {NoStop}%
\bibitem [{\citenamefont {Zhou}\ \emph {et~al.}(2022)\citenamefont {Zhou}, \citenamefont {Sethi}, \citenamefont {Liu}, \citenamefont {Wang},\ and\ \citenamefont {Liu}}]{33}%
  \BibitemOpen
  \bibfield  {author} {\bibinfo {author} {\bibfnamefont {Y.}~\bibnamefont {Zhou}}, \bibinfo {author} {\bibfnamefont {G.}~\bibnamefont {Sethi}}, \bibinfo {author} {\bibfnamefont {H.}~\bibnamefont {Liu}}, \bibinfo {author} {\bibfnamefont {Z.}~\bibnamefont {Wang}},\ and\ \bibinfo {author} {\bibfnamefont {F.}~\bibnamefont {Liu}},\ }\bibfield  {title} {\bibinfo {title} {Excited quantum anomalous and spin hall effect: dissociation of flat-bands-enabled excitonic insulator state},\ }\href@noop {} {\bibfield  {journal} {\bibinfo  {journal} {Nanotechnology}\ }\textbf {\bibinfo {volume} {33}},\ \bibinfo {pages} {415001} (\bibinfo {year} {2022})}\BibitemShut {NoStop}%
\bibitem [{\citenamefont {Zhou}\ \emph {et~al.}(2019{\natexlab{a}})\citenamefont {Zhou}, \citenamefont {Sethi}, \citenamefont {Liu}, \citenamefont {Wang},\ and\ \citenamefont {Liu}}]{34}%
  \BibitemOpen
  \bibfield  {author} {\bibinfo {author} {\bibfnamefont {Y.}~\bibnamefont {Zhou}}, \bibinfo {author} {\bibfnamefont {G.}~\bibnamefont {Sethi}}, \bibinfo {author} {\bibfnamefont {H.}~\bibnamefont {Liu}}, \bibinfo {author} {\bibfnamefont {Z.}~\bibnamefont {Wang}},\ and\ \bibinfo {author} {\bibfnamefont {F.}~\bibnamefont {Liu}},\ }\bibfield  {title} {\bibinfo {title} {Excited quantum hall effect: enantiomorphic flat bands in a yin-yang kagome lattice},\ }\href@noop {} {\bibfield  {journal} {\bibinfo  {journal} {arXiv preprint arXiv:1908.03689}\ } (\bibinfo {year} {2019}{\natexlab{a}})}\BibitemShut {NoStop}%
\bibitem [{\citenamefont {Ni}\ \emph {et~al.}(2020)\citenamefont {Ni}, \citenamefont {Zhou}, \citenamefont {Sethi},\ and\ \citenamefont {Liu}}]{35}%
  \BibitemOpen
  \bibfield  {author} {\bibinfo {author} {\bibfnamefont {X.}~\bibnamefont {Ni}}, \bibinfo {author} {\bibfnamefont {Y.}~\bibnamefont {Zhou}}, \bibinfo {author} {\bibfnamefont {G.}~\bibnamefont {Sethi}},\ and\ \bibinfo {author} {\bibfnamefont {F.}~\bibnamefont {Liu}},\ }\bibfield  {title} {\bibinfo {title} {$\pi$-orbital yin--yang kagome bands in anilato-based metal--organic frameworks},\ }\href@noop {} {\bibfield  {journal} {\bibinfo  {journal} {Physical Chemistry Chemical Physics}\ }\textbf {\bibinfo {volume} {22}},\ \bibinfo {pages} {25827} (\bibinfo {year} {2020})}\BibitemShut {NoStop}%
\bibitem [{\citenamefont {T{\"o}rm{\"a}}\ \emph {et~al.}(2021)\citenamefont {T{\"o}rm{\"a}}, \citenamefont {Peotta},\ and\ \citenamefont {Bernevig}}]{36}%
  \BibitemOpen
  \bibfield  {author} {\bibinfo {author} {\bibfnamefont {P.}~\bibnamefont {T{\"o}rm{\"a}}}, \bibinfo {author} {\bibfnamefont {S.}~\bibnamefont {Peotta}},\ and\ \bibinfo {author} {\bibfnamefont {B.~A.}\ \bibnamefont {Bernevig}},\ }\bibfield  {title} {\bibinfo {title} {Superfluidity and quantum geometry in twisted multilayer systems},\ }\href@noop {} {\bibfield  {journal} {\bibinfo  {journal} {arXiv preprint arXiv:2111.00807}\ } (\bibinfo {year} {2021})}\BibitemShut {NoStop}%
\bibitem [{\citenamefont {Herzog-Arbeitman}\ \emph {et~al.}(2022)\citenamefont {Herzog-Arbeitman}, \citenamefont {Peri}, \citenamefont {Schindler}, \citenamefont {Huber},\ and\ \citenamefont {Bernevig}}]{37}%
  \BibitemOpen
  \bibfield  {author} {\bibinfo {author} {\bibfnamefont {J.}~\bibnamefont {Herzog-Arbeitman}}, \bibinfo {author} {\bibfnamefont {V.}~\bibnamefont {Peri}}, \bibinfo {author} {\bibfnamefont {F.}~\bibnamefont {Schindler}}, \bibinfo {author} {\bibfnamefont {S.~D.}\ \bibnamefont {Huber}},\ and\ \bibinfo {author} {\bibfnamefont {B.~A.}\ \bibnamefont {Bernevig}},\ }\bibfield  {title} {\bibinfo {title} {Superfluid weight bounds from symmetry and quantum geometry in flat bands},\ }\href@noop {} {\bibfield  {journal} {\bibinfo  {journal} {Physical review letters}\ }\textbf {\bibinfo {volume} {128}},\ \bibinfo {pages} {087002} (\bibinfo {year} {2022})}\BibitemShut {NoStop}%
\bibitem [{\citenamefont {Liu}\ \emph {et~al.}(2014)\citenamefont {Liu}, \citenamefont {Liu},\ and\ \citenamefont {Wu}}]{22}%
  \BibitemOpen
  \bibfield  {author} {\bibinfo {author} {\bibfnamefont {Z.}~\bibnamefont {Liu}}, \bibinfo {author} {\bibfnamefont {F.}~\bibnamefont {Liu}},\ and\ \bibinfo {author} {\bibfnamefont {Y.-S.}\ \bibnamefont {Wu}},\ }\bibfield  {title} {\bibinfo {title} {Exotic electronic states in the world of flat bands: From theory to material},\ }\href@noop {} {\bibfield  {journal} {\bibinfo  {journal} {Chinese Physics B}\ }\textbf {\bibinfo {volume} {23}},\ \bibinfo {pages} {077308} (\bibinfo {year} {2014})}\BibitemShut {NoStop}%
\bibitem [{\citenamefont {Liu}\ \emph {et~al.}(2022)\citenamefont {Liu}, \citenamefont {Sethi}, \citenamefont {Meng},\ and\ \citenamefont {Liu}}]{23}%
  \BibitemOpen
  \bibfield  {author} {\bibinfo {author} {\bibfnamefont {H.}~\bibnamefont {Liu}}, \bibinfo {author} {\bibfnamefont {G.}~\bibnamefont {Sethi}}, \bibinfo {author} {\bibfnamefont {S.}~\bibnamefont {Meng}},\ and\ \bibinfo {author} {\bibfnamefont {F.}~\bibnamefont {Liu}},\ }\bibfield  {title} {\bibinfo {title} {Orbital design of flat bands in non-line-graph lattices via line-graph wave functions},\ }\href@noop {} {\bibfield  {journal} {\bibinfo  {journal} {Physical Review B}\ }\textbf {\bibinfo {volume} {105}},\ \bibinfo {pages} {085128} (\bibinfo {year} {2022})}\BibitemShut {NoStop}%
\bibitem [{\citenamefont {Liu}\ \emph {et~al.}(2021)\citenamefont {Liu}, \citenamefont {Meng},\ and\ \citenamefont {Liu}}]{24}%
  \BibitemOpen
  \bibfield  {author} {\bibinfo {author} {\bibfnamefont {H.}~\bibnamefont {Liu}}, \bibinfo {author} {\bibfnamefont {S.}~\bibnamefont {Meng}},\ and\ \bibinfo {author} {\bibfnamefont {F.}~\bibnamefont {Liu}},\ }\bibfield  {title} {\bibinfo {title} {Screening two-dimensional materials with topological flat bands},\ }\href@noop {} {\bibfield  {journal} {\bibinfo  {journal} {Physical Review Materials}\ }\textbf {\bibinfo {volume} {5}},\ \bibinfo {pages} {084203} (\bibinfo {year} {2021})}\BibitemShut {NoStop}%
\bibitem [{\citenamefont {Mielke}(1991)}]{25}%
  \BibitemOpen
  \bibfield  {author} {\bibinfo {author} {\bibfnamefont {A.}~\bibnamefont {Mielke}},\ }\bibfield  {title} {\bibinfo {title} {Ferromagnetism in the hubbard model on line graphs and further considerations},\ }\href@noop {} {\bibfield  {journal} {\bibinfo  {journal} {Journal of Physics A: Mathematical and General}\ }\textbf {\bibinfo {volume} {24}},\ \bibinfo {pages} {3311} (\bibinfo {year} {1991})}\BibitemShut {NoStop}%
\bibitem [{\citenamefont {Ma}\ \emph {et~al.}(2020)\citenamefont {Ma}, \citenamefont {Xu}, \citenamefont {Chiu}, \citenamefont {Regnault}, \citenamefont {Houck}, \citenamefont {Song},\ and\ \citenamefont {Bernevig}}]{26}%
  \BibitemOpen
  \bibfield  {author} {\bibinfo {author} {\bibfnamefont {D.-S.}\ \bibnamefont {Ma}}, \bibinfo {author} {\bibfnamefont {Y.}~\bibnamefont {Xu}}, \bibinfo {author} {\bibfnamefont {C.~S.}\ \bibnamefont {Chiu}}, \bibinfo {author} {\bibfnamefont {N.}~\bibnamefont {Regnault}}, \bibinfo {author} {\bibfnamefont {A.~A.}\ \bibnamefont {Houck}}, \bibinfo {author} {\bibfnamefont {Z.}~\bibnamefont {Song}},\ and\ \bibinfo {author} {\bibfnamefont {B.~A.}\ \bibnamefont {Bernevig}},\ }\bibfield  {title} {\bibinfo {title} {Spin-orbit-induced topological flat bands in line and split graphs of bipartite lattices},\ }\href@noop {} {\bibfield  {journal} {\bibinfo  {journal} {Physical review letters}\ }\textbf {\bibinfo {volume} {125}},\ \bibinfo {pages} {266403} (\bibinfo {year} {2020})}\BibitemShut {NoStop}%
\bibitem [{\citenamefont {Cvetkovic}\ \emph {et~al.}(2004)\citenamefont {Cvetkovic}, \citenamefont {Rowlinson},\ and\ \citenamefont {Simic}}]{27}%
  \BibitemOpen
  \bibfield  {author} {\bibinfo {author} {\bibfnamefont {D.}~\bibnamefont {Cvetkovic}}, \bibinfo {author} {\bibfnamefont {P.}~\bibnamefont {Rowlinson}},\ and\ \bibinfo {author} {\bibfnamefont {S.}~\bibnamefont {Simic}},\ }\href@noop {} {\emph {\bibinfo {title} {Spectral generalizations of line graphs: On graphs with least eigenvalue-2}}},\ Vol.\ \bibinfo {volume} {314}\ (\bibinfo  {publisher} {Cambridge University Press},\ \bibinfo {year} {2004})\BibitemShut {NoStop}%
\bibitem [{\citenamefont {Koll{\'a}r}\ \emph {et~al.}(2020)\citenamefont {Koll{\'a}r}, \citenamefont {Fitzpatrick}, \citenamefont {Sarnak},\ and\ \citenamefont {Houck}}]{28}%
  \BibitemOpen
  \bibfield  {author} {\bibinfo {author} {\bibfnamefont {A.~J.}\ \bibnamefont {Koll{\'a}r}}, \bibinfo {author} {\bibfnamefont {M.}~\bibnamefont {Fitzpatrick}}, \bibinfo {author} {\bibfnamefont {P.}~\bibnamefont {Sarnak}},\ and\ \bibinfo {author} {\bibfnamefont {A.~A.}\ \bibnamefont {Houck}},\ }\bibfield  {title} {\bibinfo {title} {Line-graph lattices: Euclidean and non-euclidean flat bands, and implementations in circuit quantum electrodynamics},\ }\href@noop {} {\bibfield  {journal} {\bibinfo  {journal} {Communications in Mathematical Physics}\ }\textbf {\bibinfo {volume} {376}},\ \bibinfo {pages} {1909} (\bibinfo {year} {2020})}\BibitemShut {NoStop}%
\bibitem [{\citenamefont {Brent~West}(2001)}]{38}%
  \BibitemOpen
  \bibfield  {author} {\bibinfo {author} {\bibfnamefont {D.}~\bibnamefont {Brent~West}},\ }\href@noop {} {\emph {\bibinfo {title} {Introduction to Graph Theory, vol. 2}}}\ (\bibinfo  {publisher} {Upper Saddle River, NJ: Prentice Hall},\ \bibinfo {year} {2001})\BibitemShut {NoStop}%
\bibitem [{46()}]{46}%
  \BibitemOpen
  \href@noop {} {\bibinfo {title} {See supplementary material where we have included discussion regarding the relationship between incidence matrices and square-root {TB} models, explicit {TB} hamiltonians for the cases studied in the manuscript, details about the wilson loop calculations, and examples of {BDC} construction using {2D/3D} line-graph lattices}}\BibitemShut {NoStop}%
\bibitem [{\citenamefont {Cvetkovic}\ \emph {et~al.}(1980)\citenamefont {Cvetkovic}, \citenamefont {Doob},\ and\ \citenamefont {Sachs}}]{59}%
  \BibitemOpen
  \bibfield  {author} {\bibinfo {author} {\bibfnamefont {D.}~\bibnamefont {Cvetkovic}}, \bibinfo {author} {\bibfnamefont {M.}~\bibnamefont {Doob}},\ and\ \bibinfo {author} {\bibfnamefont {H.}~\bibnamefont {Sachs}},\ }\href@noop {} {\emph {\bibinfo {title} {Spectra of graphs. Theory and application}}}\ (\bibinfo  {publisher} {Academic Press},\ \bibinfo {year} {1980})\BibitemShut {NoStop}%
\bibitem [{\citenamefont {Arkinstall}\ \emph {et~al.}(2017)\citenamefont {Arkinstall}, \citenamefont {Teimourpour}, \citenamefont {Feng}, \citenamefont {El-Ganainy},\ and\ \citenamefont {Schomerus}}]{61}%
  \BibitemOpen
  \bibfield  {author} {\bibinfo {author} {\bibfnamefont {J.}~\bibnamefont {Arkinstall}}, \bibinfo {author} {\bibfnamefont {M.}~\bibnamefont {Teimourpour}}, \bibinfo {author} {\bibfnamefont {L.}~\bibnamefont {Feng}}, \bibinfo {author} {\bibfnamefont {R.}~\bibnamefont {El-Ganainy}},\ and\ \bibinfo {author} {\bibfnamefont {H.}~\bibnamefont {Schomerus}},\ }\bibfield  {title} {\bibinfo {title} {Topological tight-binding models from nontrivial square roots},\ }\href@noop {} {\bibfield  {journal} {\bibinfo  {journal} {Physical Review B}\ }\textbf {\bibinfo {volume} {95}},\ \bibinfo {pages} {165109} (\bibinfo {year} {2017})}\BibitemShut {NoStop}%
\bibitem [{\citenamefont {Mizoguchi}\ \emph {et~al.}(2021)\citenamefont {Mizoguchi}, \citenamefont {Yoshida},\ and\ \citenamefont {Hatsugai}}]{62}%
  \BibitemOpen
  \bibfield  {author} {\bibinfo {author} {\bibfnamefont {T.}~\bibnamefont {Mizoguchi}}, \bibinfo {author} {\bibfnamefont {T.}~\bibnamefont {Yoshida}},\ and\ \bibinfo {author} {\bibfnamefont {Y.}~\bibnamefont {Hatsugai}},\ }\bibfield  {title} {\bibinfo {title} {Square-root topological semimetals},\ }\href@noop {} {\bibfield  {journal} {\bibinfo  {journal} {Physical Review B}\ }\textbf {\bibinfo {volume} {103}},\ \bibinfo {pages} {045136} (\bibinfo {year} {2021})}\BibitemShut {NoStop}%
\bibitem [{\citenamefont {Song}\ \emph {et~al.}(2020)\citenamefont {Song}, \citenamefont {Elcoro},\ and\ \citenamefont {Bernevig}}]{58}%
  \BibitemOpen
  \bibfield  {author} {\bibinfo {author} {\bibfnamefont {Z.-D.}\ \bibnamefont {Song}}, \bibinfo {author} {\bibfnamefont {L.}~\bibnamefont {Elcoro}},\ and\ \bibinfo {author} {\bibfnamefont {B.~A.}\ \bibnamefont {Bernevig}},\ }\bibfield  {title} {\bibinfo {title} {Twisted bulk-boundary correspondence of fragile topology},\ }\href@noop {} {\bibfield  {journal} {\bibinfo  {journal} {Science}\ }\textbf {\bibinfo {volume} {367}},\ \bibinfo {pages} {794} (\bibinfo {year} {2020})}\BibitemShut {NoStop}%
\bibitem [{\citenamefont {Alexandradinata}\ \emph {et~al.}(2014)\citenamefont {Alexandradinata}, \citenamefont {Dai},\ and\ \citenamefont {Bernevig}}]{alexandradinata2014wilson}%
  \BibitemOpen
  \bibfield  {author} {\bibinfo {author} {\bibfnamefont {A.}~\bibnamefont {Alexandradinata}}, \bibinfo {author} {\bibfnamefont {X.}~\bibnamefont {Dai}},\ and\ \bibinfo {author} {\bibfnamefont {B.~A.}\ \bibnamefont {Bernevig}},\ }\bibfield  {title} {\bibinfo {title} {Wilson-loop characterization of inversion-symmetric topological insulators},\ }\href@noop {} {\bibfield  {journal} {\bibinfo  {journal} {Physical Review B}\ }\textbf {\bibinfo {volume} {89}},\ \bibinfo {pages} {155114} (\bibinfo {year} {2014})}\BibitemShut {NoStop}%
\bibitem [{\citenamefont {Zhou}\ \emph {et~al.}(2019{\natexlab{b}})\citenamefont {Zhou}, \citenamefont {Jin}, \citenamefont {Huang}, \citenamefont {Wang},\ and\ \citenamefont {Liu}}]{zhou2019weyl}%
  \BibitemOpen
  \bibfield  {author} {\bibinfo {author} {\bibfnamefont {Y.}~\bibnamefont {Zhou}}, \bibinfo {author} {\bibfnamefont {K.-H.}\ \bibnamefont {Jin}}, \bibinfo {author} {\bibfnamefont {H.}~\bibnamefont {Huang}}, \bibinfo {author} {\bibfnamefont {Z.}~\bibnamefont {Wang}},\ and\ \bibinfo {author} {\bibfnamefont {F.}~\bibnamefont {Liu}},\ }\bibfield  {title} {\bibinfo {title} {Weyl points created by a three-dimensional flat band},\ }\href@noop {} {\bibfield  {journal} {\bibinfo  {journal} {Physical Review B}\ }\textbf {\bibinfo {volume} {99}},\ \bibinfo {pages} {201105} (\bibinfo {year} {2019}{\natexlab{b}})}\BibitemShut {NoStop}%
\bibitem [{\citenamefont {Lieb}(1989)}]{lieb1989two}%
  \BibitemOpen
  \bibfield  {author} {\bibinfo {author} {\bibfnamefont {E.~H.}\ \bibnamefont {Lieb}},\ }\bibfield  {title} {\bibinfo {title} {Two theorems on the hubbard model},\ }\href@noop {} {\bibfield  {journal} {\bibinfo  {journal} {Physical review letters}\ }\textbf {\bibinfo {volume} {62}},\ \bibinfo {pages} {1201} (\bibinfo {year} {1989})}\BibitemShut {NoStop}%
\bibitem [{\citenamefont {Ezawa}(2020)}]{ezawa2020systematic}%
  \BibitemOpen
  \bibfield  {author} {\bibinfo {author} {\bibfnamefont {M.}~\bibnamefont {Ezawa}},\ }\bibfield  {title} {\bibinfo {title} {Systematic construction of square-root topological insulators and superconductors},\ }\href@noop {} {\bibfield  {journal} {\bibinfo  {journal} {Physical Review Research}\ }\textbf {\bibinfo {volume} {2}},\ \bibinfo {pages} {033397} (\bibinfo {year} {2020})}\BibitemShut {NoStop}%
\bibitem [{\citenamefont {Chiu}\ \emph {et~al.}(2020)\citenamefont {Chiu}, \citenamefont {Ma}, \citenamefont {Song}, \citenamefont {Bernevig},\ and\ \citenamefont {Houck}}]{39}%
  \BibitemOpen
  \bibfield  {author} {\bibinfo {author} {\bibfnamefont {C.~S.}\ \bibnamefont {Chiu}}, \bibinfo {author} {\bibfnamefont {D.-S.}\ \bibnamefont {Ma}}, \bibinfo {author} {\bibfnamefont {Z.-D.}\ \bibnamefont {Song}}, \bibinfo {author} {\bibfnamefont {B.~A.}\ \bibnamefont {Bernevig}},\ and\ \bibinfo {author} {\bibfnamefont {A.~A.}\ \bibnamefont {Houck}},\ }\bibfield  {title} {\bibinfo {title} {Fragile topology in line-graph lattices with two, three, or four gapped flat bands},\ }\href@noop {} {\bibfield  {journal} {\bibinfo  {journal} {Physical Review Research}\ }\textbf {\bibinfo {volume} {2}},\ \bibinfo {pages} {043414} (\bibinfo {year} {2020})}\BibitemShut {NoStop}%
\bibitem [{\citenamefont {Bergman}\ \emph {et~al.}(2008)\citenamefont {Bergman}, \citenamefont {Wu},\ and\ \citenamefont {Balents}}]{40}%
  \BibitemOpen
  \bibfield  {author} {\bibinfo {author} {\bibfnamefont {D.~L.}\ \bibnamefont {Bergman}}, \bibinfo {author} {\bibfnamefont {C.}~\bibnamefont {Wu}},\ and\ \bibinfo {author} {\bibfnamefont {L.}~\bibnamefont {Balents}},\ }\bibfield  {title} {\bibinfo {title} {Band touching from real-space topology in frustrated hopping models},\ }\href@noop {} {\bibfield  {journal} {\bibinfo  {journal} {Physical Review B}\ }\textbf {\bibinfo {volume} {78}},\ \bibinfo {pages} {125104} (\bibinfo {year} {2008})}\BibitemShut {NoStop}%
\bibitem [{\citenamefont {Sunada}(2012)}]{41}%
  \BibitemOpen
  \bibfield  {author} {\bibinfo {author} {\bibfnamefont {T.}~\bibnamefont {Sunada}},\ }\href@noop {} {\emph {\bibinfo {title} {Topological crystallography: with a view towards discrete geometric analysis}}},\ Vol.~\bibinfo {volume} {6}\ (\bibinfo  {publisher} {Springer Science \& Business Media},\ \bibinfo {year} {2012})\BibitemShut {NoStop}%
\bibitem [{\citenamefont {Asratian}\ \emph {et~al.}(1998)\citenamefont {Asratian}, \citenamefont {Denley},\ and\ \citenamefont {H{\"a}ggkvist}}]{42}%
  \BibitemOpen
  \bibfield  {author} {\bibinfo {author} {\bibfnamefont {A.~S.}\ \bibnamefont {Asratian}}, \bibinfo {author} {\bibfnamefont {T.~M.}\ \bibnamefont {Denley}},\ and\ \bibinfo {author} {\bibfnamefont {R.}~\bibnamefont {H{\"a}ggkvist}},\ }\href@noop {} {\emph {\bibinfo {title} {Bipartite graphs and their applications}}},\ Vol.\ \bibinfo {volume} {131}\ (\bibinfo  {publisher} {Cambridge university press},\ \bibinfo {year} {1998})\BibitemShut {NoStop}%
\bibitem [{\citenamefont {Biggs}(1993)}]{43}%
  \BibitemOpen
  \bibfield  {author} {\bibinfo {author} {\bibfnamefont {N.}~\bibnamefont {Biggs}},\ }\href@noop {} {\emph {\bibinfo {title} {Algebraic graph theory}}},\ \bibinfo {number} {67}\ (\bibinfo  {publisher} {Cambridge university press},\ \bibinfo {year} {1993})\BibitemShut {NoStop}%
\bibitem [{\citenamefont {Zhou}\ \emph {et~al.}(2020)\citenamefont {Zhou}, \citenamefont {Sethi}, \citenamefont {Zhang}, \citenamefont {Ni},\ and\ \citenamefont {Liu}}]{44}%
  \BibitemOpen
  \bibfield  {author} {\bibinfo {author} {\bibfnamefont {Y.}~\bibnamefont {Zhou}}, \bibinfo {author} {\bibfnamefont {G.}~\bibnamefont {Sethi}}, \bibinfo {author} {\bibfnamefont {C.}~\bibnamefont {Zhang}}, \bibinfo {author} {\bibfnamefont {X.}~\bibnamefont {Ni}},\ and\ \bibinfo {author} {\bibfnamefont {F.}~\bibnamefont {Liu}},\ }\bibfield  {title} {\bibinfo {title} {Giant intrinsic circular dichroism of enantiomorphic flat chern bands and flatband devices},\ }\href@noop {} {\bibfield  {journal} {\bibinfo  {journal} {Physical Review B}\ }\textbf {\bibinfo {volume} {102}},\ \bibinfo {pages} {125115} (\bibinfo {year} {2020})}\BibitemShut {NoStop}%
\bibitem [{\citenamefont {Peri}\ \emph {et~al.}(2021)\citenamefont {Peri}, \citenamefont {Song}, \citenamefont {Bernevig},\ and\ \citenamefont {Huber}}]{45}%
  \BibitemOpen
  \bibfield  {author} {\bibinfo {author} {\bibfnamefont {V.}~\bibnamefont {Peri}}, \bibinfo {author} {\bibfnamefont {Z.-D.}\ \bibnamefont {Song}}, \bibinfo {author} {\bibfnamefont {B.~A.}\ \bibnamefont {Bernevig}},\ and\ \bibinfo {author} {\bibfnamefont {S.~D.}\ \bibnamefont {Huber}},\ }\bibfield  {title} {\bibinfo {title} {Fragile topology and flat-band superconductivity in the strong-coupling regime},\ }\href@noop {} {\bibfield  {journal} {\bibinfo  {journal} {Physical review letters}\ }\textbf {\bibinfo {volume} {126}},\ \bibinfo {pages} {027002} (\bibinfo {year} {2021})}\BibitemShut {NoStop}%
\bibitem [{\citenamefont {Po}\ \emph {et~al.}(2019)\citenamefont {Po}, \citenamefont {Zou}, \citenamefont {Senthil},\ and\ \citenamefont {Vishwanath}}]{47}%
  \BibitemOpen
  \bibfield  {author} {\bibinfo {author} {\bibfnamefont {H.~C.}\ \bibnamefont {Po}}, \bibinfo {author} {\bibfnamefont {L.}~\bibnamefont {Zou}}, \bibinfo {author} {\bibfnamefont {T.}~\bibnamefont {Senthil}},\ and\ \bibinfo {author} {\bibfnamefont {A.}~\bibnamefont {Vishwanath}},\ }\bibfield  {title} {\bibinfo {title} {Faithful tight-binding models and fragile topology of magic-angle bilayer graphene},\ }\href@noop {} {\bibfield  {journal} {\bibinfo  {journal} {Physical Review B}\ }\textbf {\bibinfo {volume} {99}},\ \bibinfo {pages} {195455} (\bibinfo {year} {2019})}\BibitemShut {NoStop}%
\bibitem [{\citenamefont {Po}\ \emph {et~al.}(2018)\citenamefont {Po}, \citenamefont {Watanabe},\ and\ \citenamefont {Vishwanath}}]{48}%
  \BibitemOpen
  \bibfield  {author} {\bibinfo {author} {\bibfnamefont {H.~C.}\ \bibnamefont {Po}}, \bibinfo {author} {\bibfnamefont {H.}~\bibnamefont {Watanabe}},\ and\ \bibinfo {author} {\bibfnamefont {A.}~\bibnamefont {Vishwanath}},\ }\bibfield  {title} {\bibinfo {title} {Fragile topology and wannier obstructions},\ }\href@noop {} {\bibfield  {journal} {\bibinfo  {journal} {Physical review letters}\ }\textbf {\bibinfo {volume} {121}},\ \bibinfo {pages} {126402} (\bibinfo {year} {2018})}\BibitemShut {NoStop}%
\bibitem [{\citenamefont {Bradlyn}\ \emph {et~al.}(2019)\citenamefont {Bradlyn}, \citenamefont {Wang}, \citenamefont {Cano},\ and\ \citenamefont {Bernevig}}]{49}%
  \BibitemOpen
  \bibfield  {author} {\bibinfo {author} {\bibfnamefont {B.}~\bibnamefont {Bradlyn}}, \bibinfo {author} {\bibfnamefont {Z.}~\bibnamefont {Wang}}, \bibinfo {author} {\bibfnamefont {J.}~\bibnamefont {Cano}},\ and\ \bibinfo {author} {\bibfnamefont {B.~A.}\ \bibnamefont {Bernevig}},\ }\bibfield  {title} {\bibinfo {title} {Disconnected elementary band representations, fragile topology, and wilson loops as topological indices: An example on the triangular lattice},\ }\href@noop {} {\bibfield  {journal} {\bibinfo  {journal} {Physical Review B}\ }\textbf {\bibinfo {volume} {99}},\ \bibinfo {pages} {045140} (\bibinfo {year} {2019})}\BibitemShut {NoStop}%
\bibitem [{\citenamefont {Ahn}\ \emph {et~al.}(2019)\citenamefont {Ahn}, \citenamefont {Park},\ and\ \citenamefont {Yang}}]{50}%
  \BibitemOpen
  \bibfield  {author} {\bibinfo {author} {\bibfnamefont {J.}~\bibnamefont {Ahn}}, \bibinfo {author} {\bibfnamefont {S.}~\bibnamefont {Park}},\ and\ \bibinfo {author} {\bibfnamefont {B.-J.}\ \bibnamefont {Yang}},\ }\bibfield  {title} {\bibinfo {title} {Failure of nielsen-ninomiya theorem and fragile topology in two-dimensional systems with space-time inversion symmetry: application to twisted bilayer graphene at magic angle},\ }\href@noop {} {\bibfield  {journal} {\bibinfo  {journal} {Physical Review X}\ }\textbf {\bibinfo {volume} {9}},\ \bibinfo {pages} {021013} (\bibinfo {year} {2019})}\BibitemShut {NoStop}%
\bibitem [{\citenamefont {Slater}\ and\ \citenamefont {Koster}(1954)}]{51}%
  \BibitemOpen
  \bibfield  {author} {\bibinfo {author} {\bibfnamefont {J.~C.}\ \bibnamefont {Slater}}\ and\ \bibinfo {author} {\bibfnamefont {G.~F.}\ \bibnamefont {Koster}},\ }\bibfield  {title} {\bibinfo {title} {Simplified lcao method for the periodic potential problem},\ }\href@noop {} {\bibfield  {journal} {\bibinfo  {journal} {Physical review}\ }\textbf {\bibinfo {volume} {94}},\ \bibinfo {pages} {1498} (\bibinfo {year} {1954})}\BibitemShut {NoStop}%
\bibitem [{\citenamefont {Delgado}\ \emph {et~al.}(2023)\citenamefont {Delgado}, \citenamefont {Dusold}, \citenamefont {Jiang}, \citenamefont {Cronin}, \citenamefont {Louie},\ and\ \citenamefont {Fischer}}]{52}%
  \BibitemOpen
  \bibfield  {author} {\bibinfo {author} {\bibfnamefont {A.}~\bibnamefont {Delgado}}, \bibinfo {author} {\bibfnamefont {C.}~\bibnamefont {Dusold}}, \bibinfo {author} {\bibfnamefont {J.}~\bibnamefont {Jiang}}, \bibinfo {author} {\bibfnamefont {A.}~\bibnamefont {Cronin}}, \bibinfo {author} {\bibfnamefont {S.~G.}\ \bibnamefont {Louie}},\ and\ \bibinfo {author} {\bibfnamefont {F.~R.}\ \bibnamefont {Fischer}},\ }\bibfield  {title} {\bibinfo {title} {Evidence for excitonic insulator ground state in triangulene kagome lattice},\ }\href@noop {} {\bibfield  {journal} {\bibinfo  {journal} {arXiv preprint arXiv:2301.06171}\ } (\bibinfo {year} {2023})}\BibitemShut {NoStop}%
\bibitem [{\citenamefont {Ortiz}\ \emph {et~al.}(2022)\citenamefont {Ortiz}, \citenamefont {Catarina},\ and\ \citenamefont {Fern{\'a}ndez-Rossier}}]{53}%
  \BibitemOpen
  \bibfield  {author} {\bibinfo {author} {\bibfnamefont {R.}~\bibnamefont {Ortiz}}, \bibinfo {author} {\bibfnamefont {G.}~\bibnamefont {Catarina}},\ and\ \bibinfo {author} {\bibfnamefont {J.}~\bibnamefont {Fern{\'a}ndez-Rossier}},\ }\bibfield  {title} {\bibinfo {title} {Theory of triangulene two-dimensional crystals},\ }\href@noop {} {\bibfield  {journal} {\bibinfo  {journal} {2D Materials}\ }\textbf {\bibinfo {volume} {10}},\ \bibinfo {pages} {015015} (\bibinfo {year} {2022})}\BibitemShut {NoStop}%
\bibitem [{\citenamefont {Dong}\ \emph {et~al.}(2021)\citenamefont {Dong}, \citenamefont {Zhang}, \citenamefont {Li}, \citenamefont {Wang}, \citenamefont {Yang}, \citenamefont {Rho}, \citenamefont {Wang}, \citenamefont {Grigoropoulos}, \citenamefont {Wu},\ and\ \citenamefont {Yao}}]{54}%
  \BibitemOpen
  \bibfield  {author} {\bibinfo {author} {\bibfnamefont {K.}~\bibnamefont {Dong}}, \bibinfo {author} {\bibfnamefont {T.}~\bibnamefont {Zhang}}, \bibinfo {author} {\bibfnamefont {J.}~\bibnamefont {Li}}, \bibinfo {author} {\bibfnamefont {Q.}~\bibnamefont {Wang}}, \bibinfo {author} {\bibfnamefont {F.}~\bibnamefont {Yang}}, \bibinfo {author} {\bibfnamefont {Y.}~\bibnamefont {Rho}}, \bibinfo {author} {\bibfnamefont {D.}~\bibnamefont {Wang}}, \bibinfo {author} {\bibfnamefont {C.~P.}\ \bibnamefont {Grigoropoulos}}, \bibinfo {author} {\bibfnamefont {J.}~\bibnamefont {Wu}},\ and\ \bibinfo {author} {\bibfnamefont {J.}~\bibnamefont {Yao}},\ }\bibfield  {title} {\bibinfo {title} {Flat bands in magic-angle bilayer photonic crystals at small twists},\ }\href@noop {} {\bibfield  {journal} {\bibinfo  {journal} {Physical review letters}\ }\textbf {\bibinfo {volume} {126}},\ \bibinfo {pages} {223601} (\bibinfo {year} {2021})}\BibitemShut {NoStop}%
\bibitem [{\citenamefont {Leykam}\ \emph {et~al.}(2018)\citenamefont {Leykam}, \citenamefont {Andreanov},\ and\ \citenamefont {Flach}}]{55}%
  \BibitemOpen
  \bibfield  {author} {\bibinfo {author} {\bibfnamefont {D.}~\bibnamefont {Leykam}}, \bibinfo {author} {\bibfnamefont {A.}~\bibnamefont {Andreanov}},\ and\ \bibinfo {author} {\bibfnamefont {S.}~\bibnamefont {Flach}},\ }\bibfield  {title} {\bibinfo {title} {Artificial flat band systems: from lattice models to experiments},\ }\href@noop {} {\bibfield  {journal} {\bibinfo  {journal} {Advances in Physics: X}\ }\textbf {\bibinfo {volume} {3}},\ \bibinfo {pages} {1473052} (\bibinfo {year} {2018})}\BibitemShut {NoStop}%
\bibitem [{\citenamefont {Apaja}\ \emph {et~al.}(2010)\citenamefont {Apaja}, \citenamefont {Hyrk{\"a}s},\ and\ \citenamefont {Manninen}}]{56}%
  \BibitemOpen
  \bibfield  {author} {\bibinfo {author} {\bibfnamefont {V.}~\bibnamefont {Apaja}}, \bibinfo {author} {\bibfnamefont {M.}~\bibnamefont {Hyrk{\"a}s}},\ and\ \bibinfo {author} {\bibfnamefont {M.}~\bibnamefont {Manninen}},\ }\bibfield  {title} {\bibinfo {title} {Flat bands, dirac cones, and atom dynamics in an optical lattice},\ }\href@noop {} {\bibfield  {journal} {\bibinfo  {journal} {Physical Review A}\ }\textbf {\bibinfo {volume} {82}},\ \bibinfo {pages} {041402} (\bibinfo {year} {2010})}\BibitemShut {NoStop}%
\bibitem [{\citenamefont {Aoki}\ \emph {et~al.}(1996)\citenamefont {Aoki}, \citenamefont {Ando},\ and\ \citenamefont {Matsumura}}]{aoki1996hofstadter}%
  \BibitemOpen
  \bibfield  {author} {\bibinfo {author} {\bibfnamefont {H.}~\bibnamefont {Aoki}}, \bibinfo {author} {\bibfnamefont {M.}~\bibnamefont {Ando}},\ and\ \bibinfo {author} {\bibfnamefont {H.}~\bibnamefont {Matsumura}},\ }\bibfield  {title} {\bibinfo {title} {Hofstadter butterflies for flat bands},\ }\href@noop {} {\bibfield  {journal} {\bibinfo  {journal} {Physical Review B}\ }\textbf {\bibinfo {volume} {54}},\ \bibinfo {pages} {R17296} (\bibinfo {year} {1996})}\BibitemShut {NoStop}%
\bibitem [{\citenamefont {Marques}\ \emph {et~al.}(2023)\citenamefont {Marques}, \citenamefont {M{\"o}gerle}, \citenamefont {Pelegr{\'\i}}, \citenamefont {Flannigan}, \citenamefont {Dias},\ and\ \citenamefont {Daley}}]{marques2023kaleidoscopes}%
  \BibitemOpen
  \bibfield  {author} {\bibinfo {author} {\bibfnamefont {A.}~\bibnamefont {Marques}}, \bibinfo {author} {\bibfnamefont {J.}~\bibnamefont {M{\"o}gerle}}, \bibinfo {author} {\bibfnamefont {G.}~\bibnamefont {Pelegr{\'\i}}}, \bibinfo {author} {\bibfnamefont {S.}~\bibnamefont {Flannigan}}, \bibinfo {author} {\bibfnamefont {R.}~\bibnamefont {Dias}},\ and\ \bibinfo {author} {\bibfnamefont {A.}~\bibnamefont {Daley}},\ }\bibfield  {title} {\bibinfo {title} {Kaleidoscopes of hofstadter butterflies and aharonov-bohm caging from 2 n-root topology in decorated square lattices},\ }\href@noop {} {\bibfield  {journal} {\bibinfo  {journal} {Physical Review Research}\ }\textbf {\bibinfo {volume} {5}},\ \bibinfo {pages} {023110} (\bibinfo {year} {2023})}\BibitemShut {NoStop}%
\end{thebibliography}%
\end{document}